\documentclass[superscriptaddress,secnumarabic,amssymb, nobibnotes, aps, prf]{revtex4-1}
\usepackage{graphicx}
\usepackage{amsfonts}
\usepackage{amssymb}
\usepackage{amsmath}
\usepackage{psfrag}
\usepackage{natbib}
\usepackage{mathtools}
\usepackage{ulem}
\usepackage[colorlinks=true, linkcolor=blue, urlcolor=blue, citecolor=black]{hyperref}
\usepackage{caption}
\newcommand{\beqa}{\begin{eqnarray}}
\newcommand{\eeqa}{\end{eqnarray}}
\newcommand{\beq}{\begin{equation}}
\newcommand{\eeq}{\end{equation}}

\providecommand \selectlanguage [0]{\@gobble}%

\begin{document}

\title{Non-Destructive Detection of Early Drought Stress\\ via Spectral Submanifold Analysis}

\author{Sunghwan Jung}
\email{sj737@cornell.edu}
\affiliation{Biological and Environmental Engineering, Cornell University, Ithaca, NY 14853, USA}

\author{Isabella Thiel}
\thanks{These authors contributed equally to this work.}
\affiliation{Department of Mechanial Engineering, MIT, Cambridge, MA 02139, USA}

\author{Jisoo Yuk}
\thanks{These authors contributed equally to this work.}
\affiliation{Biological and Environmental Engineering, Cornell University, Ithaca, NY 14853, USA}

\author{George Haller}
\email{georgehaller@ethz.ch}
\affiliation{Institute for Mechanical Systems, ETH Zurich, ZH 8092, Switzerland}

\date{\today}

\begin{abstract}
Non-destructive early detection of water stress is critical for precision agriculture, yet current detection methods are often labor-intensive or destructive. We propose a rapid, non-invasive method based on Spectral Submanifold (SSM) reduction to analyze the nonlinear vibration dynamics of soybean plants. Unlike traditional linear modal analysis, our data-driven approach reveals a distinct transition in the dominant dynamics of the plant stem as drought progresses. Specifically, we uncover a sign reversal in the nonlinear damping coefficient within SSM-reduced models of full plants. Our finding reveals a fundamental transition in the plant's dissipative mechanics from a pressurized, fluid-rich vessel exhibiting superlinear (viscoelastic) damping to a dry, cellular solid dominated by sublinear (frictional) damping. This transition is closely correlated with the onset of physiological stress indicated by a significant decrease in the values of mature-leaf soil plant analysis development (SPAD) and colorimetric metrics. In contrast, traditional metrics, such as linear frequency and leaf wilting, do not provide consistent statistical signals of water stress. This study provides the first experimental evidence that SSM-based nonlinear model reduction can detect the changing internal state of a living organism, offering a physics-based metric for high-throughput plant phenotyping. \\

\textbf{Significant Statement:} Global water scarcity threatens agricultural productivity, which requires rapid and early detection of droughts. Current diagnostic methods are often destructive or labor-intensive. Here, we introduce a non-invasive, video-based method using nonlinear modal analysis to detect early water stress in plants. By evaluating the natural vibrations of soybeans, we uncover a fundamental mechanical transition: as plants lose internal water pressure, their dissipative mechanics shift from fluid-dominated viscoelasticity to structure-dominated friction. This physiological phase change is clearly marked by a sign reversal in the nonlinear damping coefficient of the plant, which precedes severe visible wilting. This data-driven approach provides a physics-based metric for high-throughput crop phenotyping, offering a scalable tool for precision agriculture and water management in fields.
\end{abstract}

 \maketitle

\section{Introduction}

Global water scarcity poses a critical threat to agricultural productivity \cite{unchronicle2023water,biswas2025water}, which requires the development of drought-resistant crops and efficient irrigation strategies \cite{franco2025advancements} guided by accurate measurement of water-stress. High-throughput phenotyping enables rapid and non-destructive assessments of plant physiological status in real-time \cite{sathiya2023smart,zhao2019crop}. Established water-stress diagnostics are, however,  severely limited by their practical implementation, as they are often labor-intensive and require manual, point-by-point measurements on individual leaves \cite{brodribb2017evolution, jones2004application}. Other common methodologies, such as the widely-used Scholander pressure chamber \cite{scholander1965sap}, are inherently invasive and destructive, requiring the excision of plant tissue, or the insertion of sensor needles directly into the vascular system to monitor sap flux or internal turgor \cite{burgess2001improved}. Therefore, there is a critical need for rapid, non-destructive physical metrics that can detect the early onset of turgor loss, the hydraulic precursor to systemic physiological stress.

Among the main agricultural commodities, soybean (\textit{Glycine max} L.) is one of the most economically and nutritionally significant crops worldwide \cite{pagano2016importance,oecdfao2023outlook}, serving as a primary global source of plant-based protein, edible oil, and biofuel \cite{GAONKAR201973, graham2003legumes}. Despite its agricultural importance, soybean is highly susceptible to abiotic stresses, particularly water deficit presumably due to their broad leaves. Drought stress is widely recognized as the primary limiting factor for soybean production, capable of causing yield reductions of up to 40--60\% when severe stress coincides with early developmental stages \cite{manavalan2009physiological, mutava2015understanding}. As climate change accelerates the frequency and severity of agricultural droughts, developing rapid, non-destructive methods to monitor the early onset of drought stress in soybeans is crucial for optimizing crop management.

Previous studies treating plants as linear beams (Euler-Bernoulli beams) typically only evaluate frequency and stiffness \cite{gart2015droplet,py2006frequency,delangre2008effects}. From a biomechanical perspective, a plant stem is not only a static structural element but also a complex composite material pre-stressed by internal hydrostatic pressure, known as turgor \cite{niklas1992plant,dominguez2011overview,salmen1984viscoelastic}. This turgor pressure provides the rigidity necessary for upright growth and mechanical stability. As a plant experiences a water deficit, the loss of turgor physically alters its structural mechanics. The plant system transitions from a pressurized, fluid-filled vessel to a dry cellular tube \cite{kanahama2023mechanics,kanahama2023rigidity}. % Intuitively, such a phase change in the internal material state should alter the plant's dynamic response to physical perturbations. Therefore, a distinct change in the fluttering mode is expected.
Although traditional linear modal analysis can track the fundamental mode of kinematic motions, it inherently fails to capture the complex, amplitude-dependent dynamics driven by physiological changes. Because linear models approximate these dissipative mechanics as constant, they could not resolve the physiological phase change that occurs within the stem. Therefore, detecting this distinct shift in the plant's dynamic response is necessary to move beyond linear assumptions to a rigorous nonlinear modal analysis.

The value of soil plant analysis development (SPAD) serves as a measure of the relative chlorophyll content and acts as a physiological indicator to predict a plant's capacity to recover from drought stress \cite{zegada2013photosynthetic,Hasanuzzaman2022}. Research on barley indicates a positive correlation between SPAD values of drought-stressed plants and the Drought Recovery Index, suggesting that genotypes capable of maintaining higher chlorophyll density under stress can resume photosynthesis and growth more rapidly upon rehydration \cite{Hasanuzzaman2022}. Similarly, in sorghum, drought-tolerant genotypes exhibiting high yield stability were found to maintain their SPAD status during plant stresses \cite{SinghBakala2025}. For soybeans, while drought stress reduces SPAD values, rehydration triggers rapid recovery; in some cases, the chlorophyll content can even exceed that of control plants, particularly when stress occurs during vegetative stages \cite{Dong2019}. Therefore, SPAD is regarded as a non-destructive but labor-intensive method to evaluate photosynthetic resilience and recovery potential \cite{SinghBakala2025}.

\begin{figure}
    \centering
    \includegraphics[width=1\linewidth]{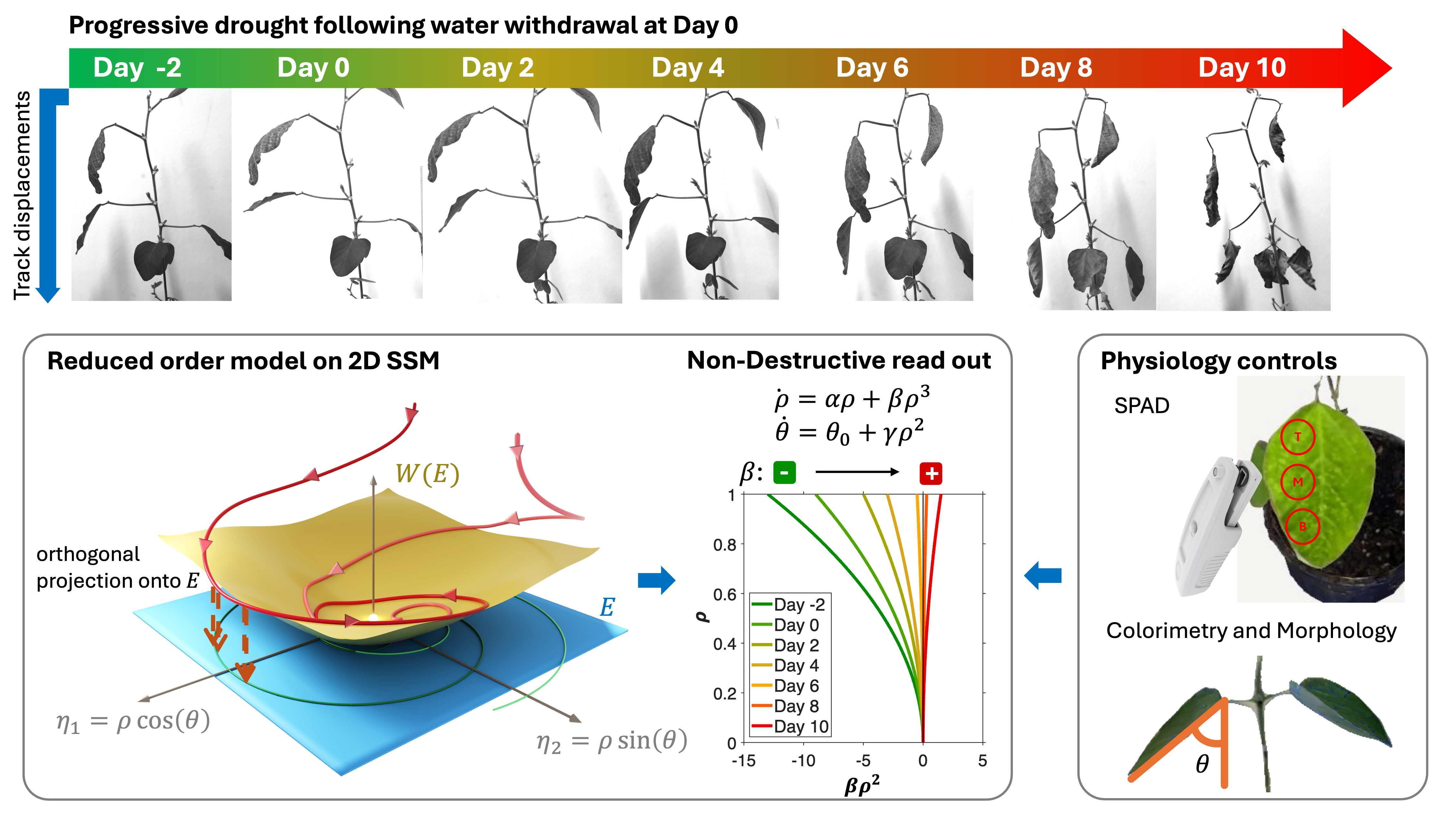}
    \caption{\textbf{Non-destructive dynamical phenotyping of progressive drought stress.} Representative soybean plant imaged from 2 days before to 10 days after watering was withheld at Day 0, showing progressive loss of turgor and visible wilting. Stem motions recorded during brief free-decay tests are reconstructed in delay coordinates and reduced onto a two-dimensional spectral submanifold  SSM $W(E)$, yielding a low-dimensional model of the dominant plant dynamics. The normal form  $\dot \rho = \alpha \rho + \beta \rho ^3, \; \dot \theta = \omega_0 + \gamma \rho^2$ separates linear decay, amplitude-dependent damping, and frequency modulation. With progressive dehydration, $\beta$ shifts from negative to near-zero or positive values, consistent with a transition from superlinear, fluid-mediated dissipation to sublinear, friction-dominated dissipation. Independent physiological benchmarks, such as SPAD, colorimetric, and morphological measurementswere not part of the dynamical inference itself, but served as external references for plant hydration and stress state to validate the SSM-derived results. Together, the workflow converts short vibration recordings into a physics-based, non-destructive readout of hydration-dependent plant mechanics.
}
    \label{fig:main_image}
\end{figure}

Free from the limitations of linear modeling and SPAD analysis, the mathematical method used in this paper to model plant dynamics from observational data is reduction to spectral submanifolds (SSMs). SSMs are ubiquitous and robust low-dimensional invariant manifolds in dynamical systems that can be used for model reduction. Specifically, a slow SSM is the unique smoothest invariant manifold that emanates from a stationary state along the state's dominant spectral subspace. Such a subspaces is the span of the observed slowest exponentially decaying (including possible exponentially growing) solutions of the linearized system at the steady  state.  If all linear solutions outside such a dominant spectral subspaces decay, then nearby trajectories of the full nonlinear system decay exponentially fast to, and synchronize with, the internal dynamics of the slow  SSM. As a consequence, the dynamics within the SSM serve as mathematically justified reduced-order models for the full nonlinear  system \cite{Haller2016,cenedese2022data}. Such SSM-reduced models have been used for system identification and control in a variety of settings, including structural vibrations of beams and shells, fluid-structure interactions, transition to turbulence, table-top gravity measurements, and the control of soft robots (see \cite{haller2025SSM} for a recent general introduction and a review of applications).
Figure ~\ref{fig:main_image} summarizes the experimental workflow: trajectory data of decaying tip displacement is recorded for plants under progressive water withdrawal. SSM-based model reduction, and independent physiological measurements are used to determine whether hydration loss leaves a distinct nonlinear mechanical signature.

In this work, we demonstrate that the dominant nonlinear vibration signature of a plant serves as a non-destructive and rapid indicator of drought stress. Our motivating hypothesis is that the loss of turgor pressure triggers a mechanical transition, switching the stem's damping mechanism from fluid-dominated viscoelasticity to structure-dominated friction. To detect this transition, we use SSM reduction to identify the lowest-dimensional dynamics organizing the observed data from videos of free plant oscillations and to uncover nonlinear backbone curves. These curves are formed by instantaneous frequency-amplitude  plots of decaying plant oscillations within slow SSMs. Studying the evolution of backbone curves in time, we uncover a distinct sign reversal in the nonlinear damping coefficient of the plant that correlates with the onset of physiological stress. Interestingly, such correlations are fundamentally nonlinear: they do not emerge for the linear stiffness and linear damping of the plant. These results promise a new, generally applicable non-destructive method for precision agriculture that provide a mechanically interpretable measure of internal physiological state.

\section{Results and Discussion}

\begin{figure}[h]
    \centering
    \includegraphics[width=.9\textwidth]{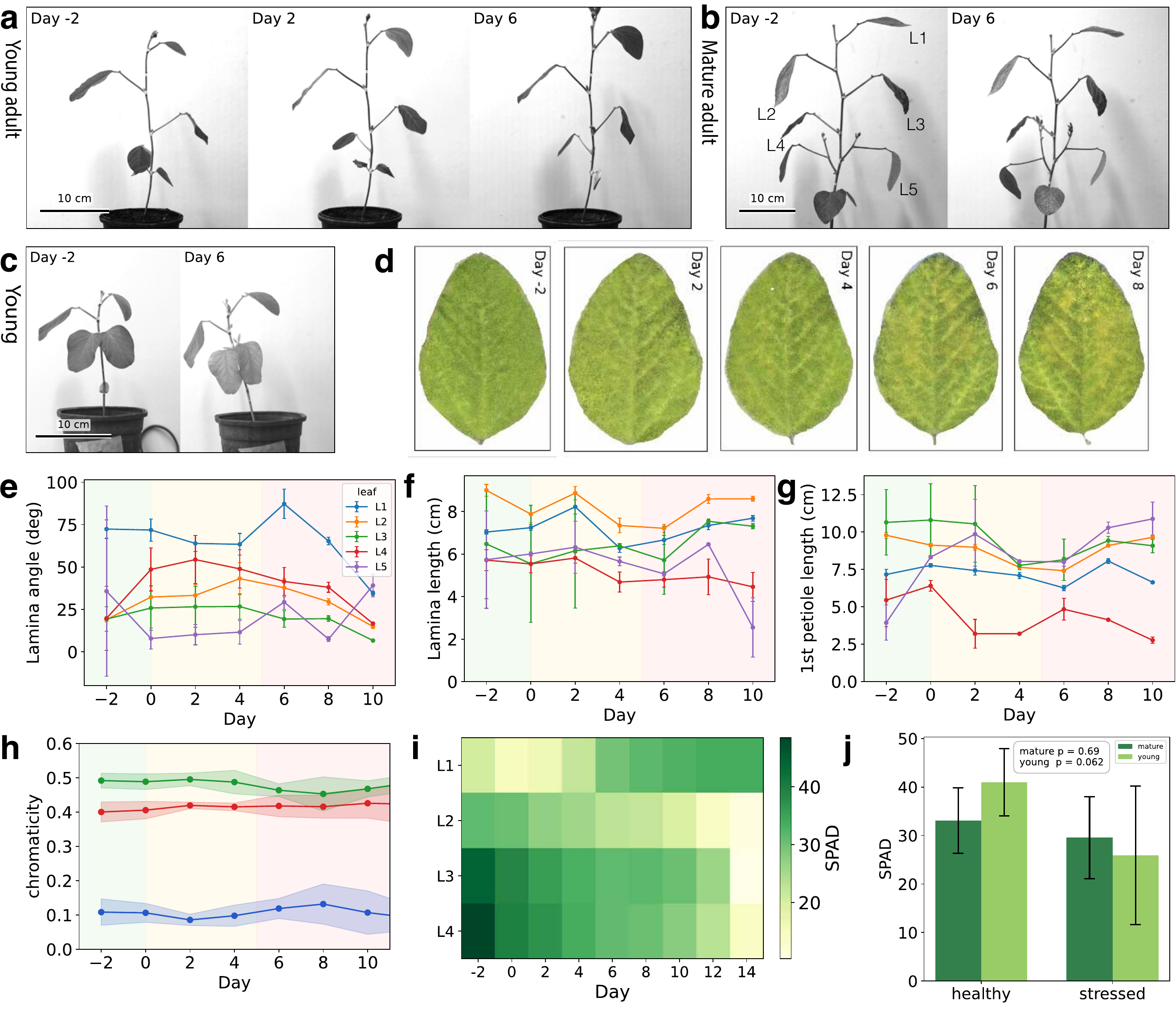}
\caption{\textbf{Morphological and physiological responses of soybean to progressive drought stress.} 
(a–c) Representative macroscopic time-lapse images of three individual soybean plants transitioning from well-watered conditions (Day -2) to water deficit (Day 6; 6 days without water). Panel c demonstrates the leaf numbering convention used for kinematic tracking, where L1 represents the youngest fully expanded leaf and L5 represents the oldest mature leaf.
(d) Isolated RGB image of a representative leaf tracking visual color from Day -2 to Day 8.
(e–g) Kinematic quantification of leaf structural parameters: (e) lamina angle relative to the vertical axis, (f) total lamina length, and (g) petiole length. Colored background spans indicate distinct experimental phases (e.g., baseline, mild stress, and water-stress). Data represent mean $\pm$ Standard Error of the Mean (SEM).
(h) Temporal changes in the mean normalized chromaticity (red, green, and blue color channels) of the isolated leaf tissue.
(i) Spatiotemporal heatmap of relative chlorophyll content (SPAD) across individual leaves (L1–L4) showing accelerated chlorophyll degradation in older foliage as the drought progresses.
(j) Paired comparison of SPAD values between mature and young leaves under healthy (Day $\le$ 0) and stressed (Day $\ge$ 5) conditions. Significance was determined using the two-sided Wilcoxon signed-rank test; young leaves exhibited a highly significant reduction in chlorophyll ($p = 0.062$), whereas mature leaves maintained relatively stable but lower baseline levels ($p = 0.69$). Error bars indicate standard deviation. }
\label{Fig1}
\end{figure}

We first performed experiments to track drought stress in \textit{Glycine max}, commonly known as soybean. The plants were monitored over a 8-15 day period, transitioning from a fully turgid state (Day -2 to Day 0) to a severe drought state (Day +5), with Day 0 marking the exact date at which watering was completely withheld. Figure~\ref{Fig1}(a-c) shows representative macroscopic time series  of three individual soybean plants as drought progressed. In Figure~\ref{Fig1}(d), visual color degradation is tracked using a 3D scan, showing parts of the tissue turning yellow as senescence sets in. Figure~\ref{Fig1}(e-g) shows the temporal evolution of the structural parameters of the leaf, including the angle of the lamina from the vertical down line, the lengths of the lamina, and that of the petiole. The data reveal a distinct two-phase structural response. Up to Day +6, the leaves altered their leaf angles without a clear trend. However, after day +6, the canopy leaves exhibited a rapid decline in leaf angle (Fig.~\ref{Fig1}(e)). This indicates a loss of turgor that physically precedes the severe morphological wilting. However, statistical analysis (detailed later in Figure~\ref{Fig4}) reveals that leaf angle alone is highly variable between individual plants and therefore is not a reliable biomarker for early detection of drought. Similarly, other physiological measures, such as the lengths of the lamina and petiole, fluctuate without a clear trend, as seen in Fig.~\ref{Fig1}(f,g). The chromaticity data quantitatively support the visual observations of non-uniform color degradation, as the drought progresses. As in Fig.~\ref{Fig1}(h), the variation in the RGB components arises because some leaves undergo chlorosis and turn yellow, while others simultaneously thicken and turn darker green.

To quantify the physiological impact of this water stress, we monitored Soil Plant Analysis Development (SPAD) values as proxies for relative chlorophyll content and overall plant health. As illustrated in the spatiotemporal heatmap (Fig.~\ref{Fig1}(i)), chlorophyll degradation was highly position-dependent. Upper, younger leaves initially had lower SPAD values that increased during the early stage of drought before stabilizing. Conversely, lower, mature leaves began with high SPAD values but experienced a significant decline under water stress. This differential response reflects a well-documented survival strategy \cite{Chenetal2015,Tanetal2023,Hasanuzzaman2022, SinghBakala2025} as plants induce early senescence in older leaves to remobilize nutrients and prioritize the survival of younger tissues. As shown in Figure~\ref{Fig1}(j), a paired comparison confirms a highly significant reduction in chlorophyll for young leaves under stress ($p=0.016$), whereas mature leaves maintained a stable but lower baseline. Generally, Days +5 to +7 appear to be the critical tipping point where broader photosynthetic metabolism begins to degrade in response to prolonged water deficit.

\begin{figure}[h]
    \centering
    \includegraphics[width=1\textwidth]{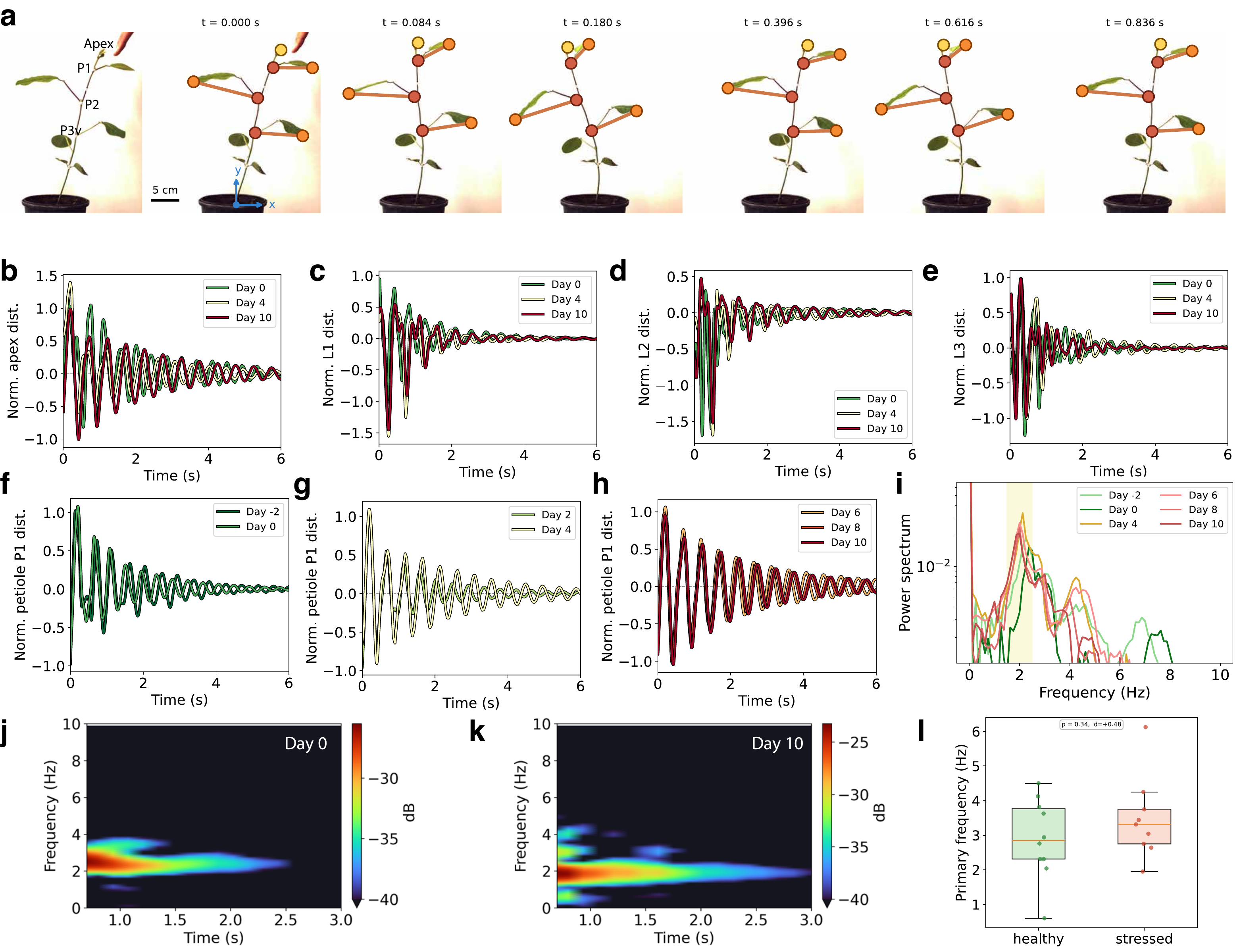}
\caption{\textbf{Kinematics and linear spectral analysis of the stem-leaf structure during progressive drought stress.} (a) Time-lapse sequence of the free-vibration "pluck" test on a healthy soybean plant. Distinct kinematic tracking points were marked along the main stem apex (yellow), petioles (P1-P3, pink), and leaf tips (orange circles). (b-e) Normalized time-series distance tracking of the (b) stem apex, and leaf tips (c) L1, (d) L2, and (e) L3, showing complex, multi-modal oscillation across selected hydration dates. (f-h) Normalized distance of the first petiole (P1) separated into (f) healthy stages (Days -2 to 0), (g) early stage (Days 2 to 4), and (g) severe stress stages (Days 6 to 10), highlighting the lengthening of the decay profile as structural stiffness decreases. (i) Power spectrum (Fast Fourier Transform, FFT) of the vibrational response, demonstrating a slight but non-significant leftward shift in the fundamental frequencies as the plant dehydrates. (j,k) Short-Time Fourier Transform spectrograms visualizing the transient frequency content during the decay period for a (j) healthy (Day 0) and (k) water-stressed (Day 10) state. (l) Box plot comparing the extracted primary linear frequency ($Hz$) of plants in healthy (Day $\le$ 0) versus water-stressed (Day $\ge$ 5) states ($p = 0.34$, $d=0.48$, paired t-test). The lack of statistical significance shows the limitation of traditional linear modal analysis in detecting early-stage turgor loss.}
\label{Fig2}
\end{figure}

To study changes in the mechanical properties of the stem-leaf structure under water stress, we performed free-vibration pluck tests (Fig.~\ref{Fig2}(a)). The resulting normalized time-series displacement data of the stem apex, leaf tips, and petioles relative to the stem base (Fig.~\ref{Fig2}(b-g)) reveal the complex, multi-modal dissipative dynamics of the system. In the healthy, fully turgid state, the vibration decays rapidly, characteristic of a highly damped structure. As drought stress progresses into severe stages, the decay profile noticeably increases in periodicity and slowly decays (as distinctly highlighted by the petiole tracking in Fig.~\ref{Fig2}(f,g,h)), indicating a loss of structural stiffness. Fast Fourier Transform (FFT) and Short-Time Fourier Transform (STFT) analyses of the decay response (Fig.~\ref{Fig2}(i-k)) show a slight leftward shift in fundamental frequency and a narrowing of spectral peaks. However, linear modal analysis which relies on assumptions of constant frequency ($f_0$), fails to resolve the transient dynamics reliably. Comparing the primary linear frequency of healthy versus stressed states yields no statistical significance ($p=0.34$, $d=0.48$, Fig.~\ref{Fig2}(l)), highlighting the limitations of linear methods.

\begin{figure}[h]
    \centering
    \includegraphics[width=1\textwidth]{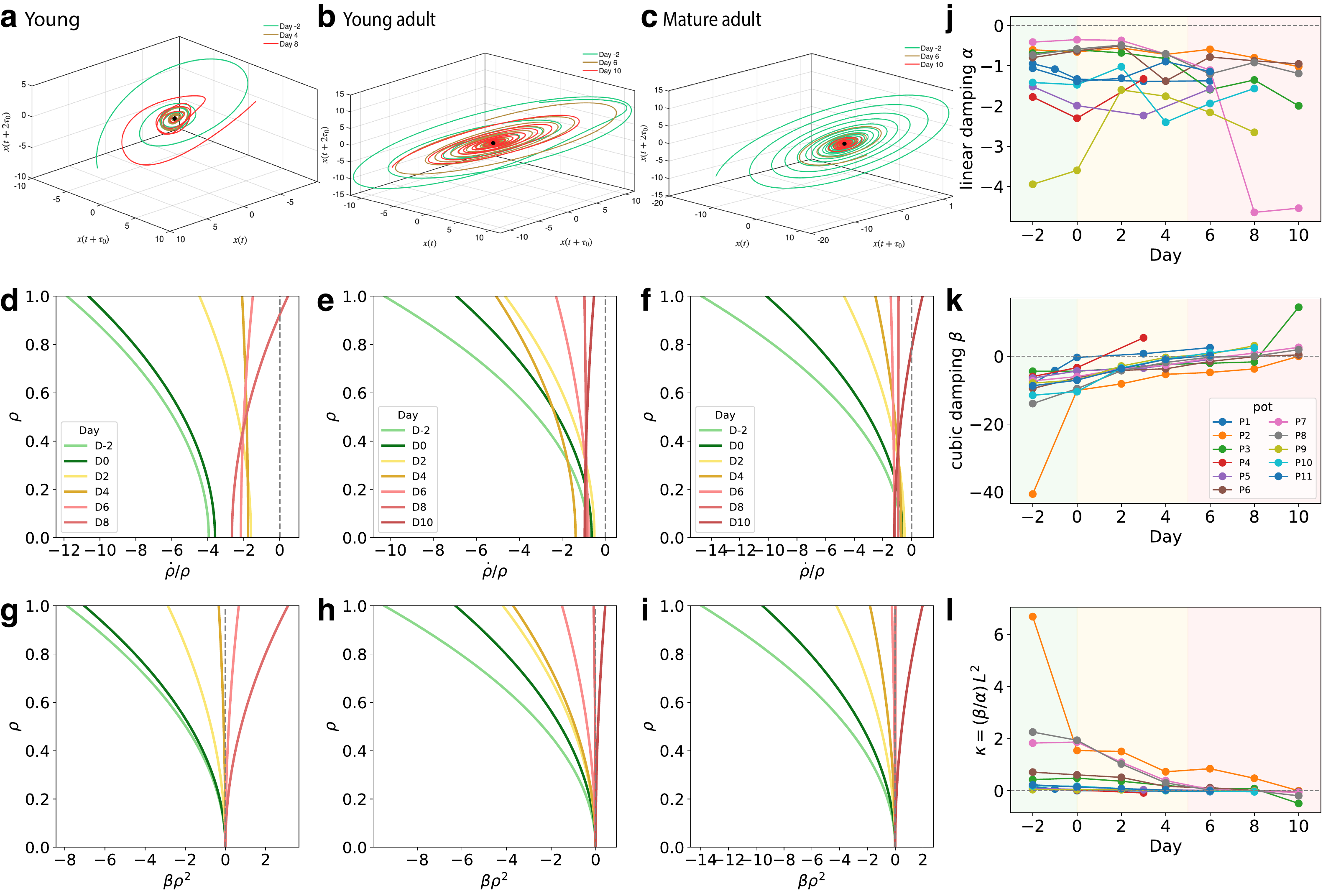}
\caption{\textbf{Nonlinear vibration dynamics and Spectral Submanifold (SSM) model reduction across hydration states.} (a-c) Delay-embedded, three-dimensional phase-space trajectories reconstructed from time-series stem displacement data of a representative plant (same plant as in Fig. \ref{Fig1}a) across three hydration states: healthy (Day -2), intermediate stress (Day +6), and water-stressed (Day +10) for three different plant development stages: (a) young, (b) young adult, and (c) mature adult. The delay embedding $(\tau = T, 2T)$ reveals the underlying invariant manifolds. (d-f) Extracted nonlinear damping backbone curves plotting the total instantaneous damping rate $\dot{\rho}/\rho$ as a function of the vibration amplitude $\rho$ for three individual plants: (d) young (Fig. \ref{Fig1}c), (e) young adult (Fig. \ref{Fig1}a), and (f) mature adult (Fig. \ref{Fig1}b). Healthy states (green) show a distinct bend toward negative values, indicating a strongly dissipative, superlinear viscoelastic response. As drought progresses (yellow to red), the curves lose curvature, reflecting a transition toward sublinear structural friction. (g-i) Isolated nonlinear damping terms $(\beta\rho^{2})$ corresponding to the backbone curves in panels d-f. (j) Evolution of the linear damping coefficient $(\alpha)$ over the experimental timeline for all tested plants $(n=10)$. (k) Evolution of the raw nonlinear damping coefficient $(\beta)$, displaying a consistent transition from negative (superlinear) to positive (sublinear) values. (l) Evolution of the dimensionless nonlinear damping ratio $(\mathcal{K} = (\beta/\alpha)L^2)$. The sign reversal of $\mathcal{K}$ ($\mathcal{K} = 0$, dashed line) serves as a robust, geometry-independent mechanical bifurcation threshold indicating severe drought stress. }  
    \label{Fig3}
\end{figure}

\begin{figure}[h]
    \centering
    \includegraphics[width=1\textwidth]{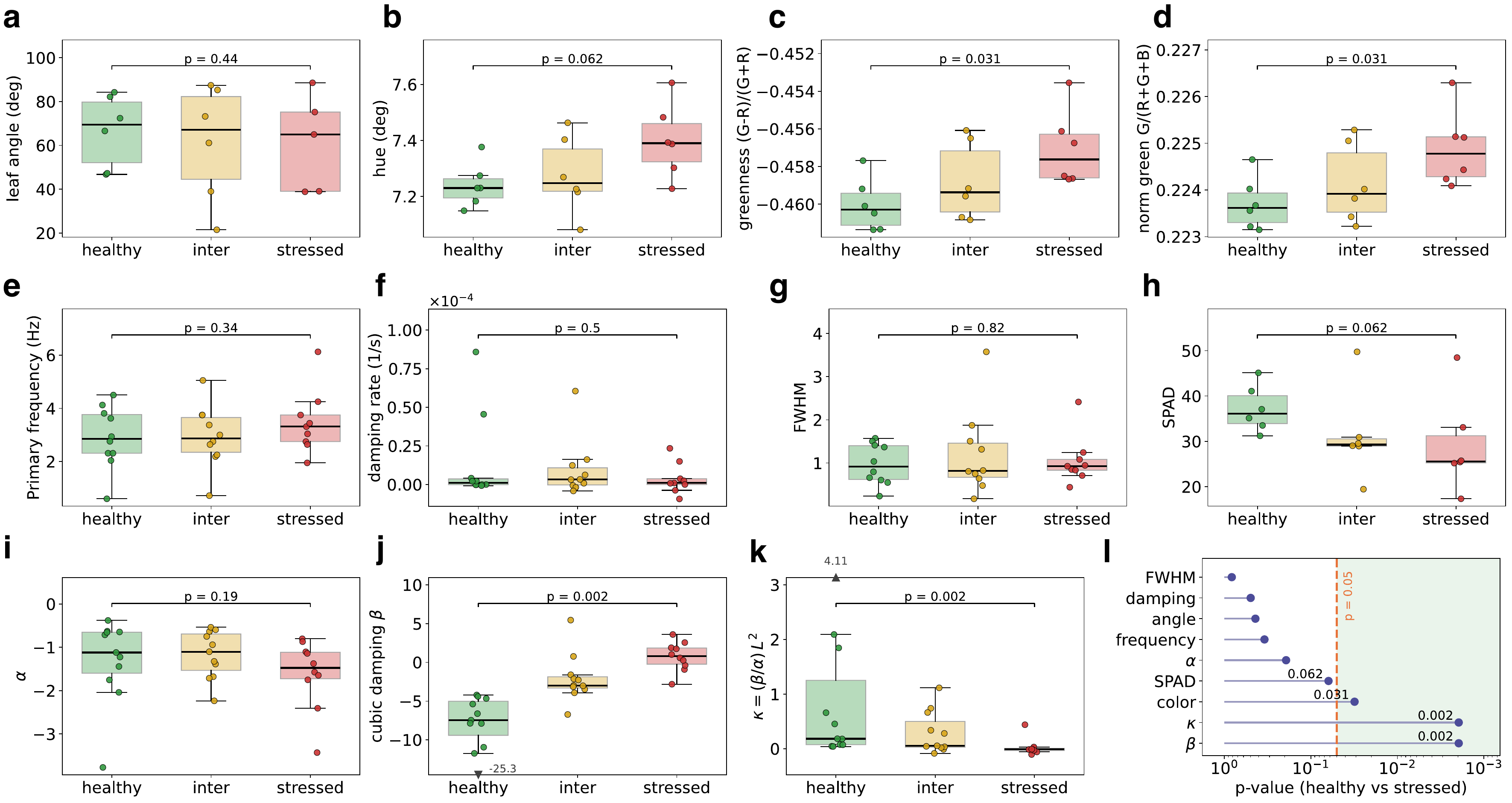}
    \caption{\textbf{Statistical comparison of morphological, pigmentary, and kinematic drought stress indicators.} Box plots track the progression of plant metrics across three defined hydration states: healthy (Days -2 to 0), intermediate stress (Days 1 to 5), and dried/stressed (Days $\ge$ 6). (a) Average leaf angle. (b-d) Pigmentary changes derived from RGB imaging of the leaves: (b) hue angle, (c) greenness index $\frac{G-R}{G+R}$, and (d) normalized green channel $\frac{G}{R+G+B}$. (e-g) Traditional linear modal analysis parameters: (e) fundamental frequency ($f_0$), (f) linear damping rate, and (g) Full Width at Half Maximum (FWHM) of the spectral peak. (h) Mean SPAD values (relative chlorophyll content) showing a marked decline from the healthy state. (i-k) Damping parameters extracted via Spectral Submanifold (SSM) reduction: (i) linear damping coefficient ($\alpha$), (j) cubic nonlinear damping coefficient ($\beta$), and (k) the dimensionless nonlinear damping ratio ($\mathcal{K}$). Both $\beta$ and $\mathcal{K}$ exhibit a highly significant sign reversal ($p=0.002$), transitioning from negative to positive values as the plant dehydrates. (l) Summary forest plot ranking the $p$-values (two-sided Wilcoxon signed-rank test, healthy vs. stressed states) of all evaluated metrics. The dashed red line indicates the standard significance threshold ($p = 0.05$). The nonlinear damping parameters ($\beta$ and $\mathcal{K}$) significantly outperform traditional kinematic metrics (linear damping, $f_0$, FWHM) and visual morphology (leaf angle), rivaling direct colorimetric metrics for non-destructive drought detection. Data points represent individual plant averages.}  
    \label{Fig4}
\end{figure}

To extract the nonlinear dynamics from the observed fluttering motions, we applied SSM reduction \cite{Haller2016,cenedese2022data} to restrict the nominally infinite-dimensional plant dynamics to a two-dimensional (2D), attracting invariant manifold emanating from the undisturbed equilibrium point of the plant. The internal dynamics within this 2D SSM serves as a low-dimensional, mathematically exact reduced model with which all decaying plant oscillations synchronize exponentially fast.   
Although the delay-embedded phase-space trajectories of P1 distance (Fig.~\ref{Fig3}(ac)) show visually distinct changes in the oscillation patterns, the fundamental mechanical transition remains unclear within these complex orbits. To explicitly reveal this hidden transition, we used an SSM-reduced model to extract the backbone curves (Fig.~\ref{Fig3}(d-f)). This characterizes how the instantaneous damping rate ($\dot{\rho}/\rho$) evolves as a function of the displacement amplitude ($\rho$). In healthy, turgid plants, the damping curves exhibit a distinct leftward bend, showing that the effective damping rate becomes more negative (i.e., highly dissipative) at larger vibration amplitudes. As drought stress advances, this line curvature straightens out, indicating a clear transition in linear dissipation. We isolate these nonlinear damping terms (Fig.~\ref{Fig3}(g-i)) and evaluate the coefficients governing the amplitude-dependent decay rate governed by the amplitude component of a universal normal form 
\begin{equation} \label{eq:decay}
    \frac{1}{\rho} \frac{d \rho}{dt}  = \alpha + {\beta}\, \rho^2  \,
\end{equation}
of the SSM-reduced dynamics. The linear term $\alpha$ in this normal form captures baseline linear dissipation near equilibrium, while the nonlinear term $\beta \rho^2$ represents nonlinear dissipation that governs the curvature of the backbone. To capture the relative contribution of this nonlinear dissipation, we define a dimensionless nonlinear damping ratio, denoted as $\mathcal{K}$, as 
\begin{equation}
    \mathcal{K} = \frac{\beta}{\alpha} \, L^2,
\end{equation}
where $L$ is the distance from the soil to the base of the leaf. 

The main finding of this study is  a distinct nonlinearity transition in the plant's vibration dynamics that signal water stress. This transition is manifested by the nonlinear damping coefficient $\beta$ approaching zero or even exhibiting a complete sign reversal (Fig.~\ref{Fig3}(k)), which drives the corresponding shift in $\mathcal{K}$. Figure~\ref{Fig3}(l) plots the evolution of $\mathcal{K}$ over the period of the drought experiment, showing a clear transition where $\mathcal{K}$ changes sign or gets close to zero between the healthy state and the irreversibly dried state. This nonlinear transition reflects the loss of water pressure in the xylem, fundamentally altering the rigidity and dissipative mechanics of the plant. 

Here, we interpret the amplitude dependence, which serves as a proxy for velocity dependence since the vibration frequency remains almost constant. In turgid plants (Days -2 to 0), the stem acts as a composite structure pre-stressed by internal water pressure. Mechanically, this creates a viscoelastic system where energy dissipation is driven by fluid motions within tiny xylem and cellular matrix or tubular structures. As the vibration amplitude increases, the internal fluid experiences increased dissipative shear forces proportional to velocity to a certain power. This results in hardening or superlinear damping, where the structure resists vibration more intensely at higher velocities and amplitudes. That presumably benefits plants by preventing catastrophic failure during high-amplitude vibrations (e.g., strong winds). As drought progresses (Day +5 or later), the loss of turgor pressure and water content in the tissues reduces the internal pre-stress and increases the contact areas of dry tissues without fluid lubricants. The stem transitions to a behavior dominated by the structural mechanics of dry fibers and cell walls. In this state, energy dissipation is largely driven by internal dry friction. Unlike viscous damping, dry friction is independent of the magnitude of velocity or amplitude. Consequently, even though the vibration amplitude (and thus velocity) increases, the effective damping ratio gets close to zero. This softening or sublinear damping signature is characteristic of dry, hysteresis-dominated structures, often modeled as a stick-slip mechanism in composite materials.

We might also consider the impact of air drag on our findings. However, because the external air resistance acting on the leaf's surface area remains relatively constant across both healthy and dried states, we can rule out air resistance as the primary driver of the  nonlinearity transition we have observed. Therefore, this transition is fundamentally an internal material property change caused by the loss of hydraulic conductivity and viscoelasticity.

To demonstrate the robustness of our nonlinear approach, we performed a comprehensive statistical comparison of all morphological, pigmentary, and kinematic indicators across hydration states (Fig.~\ref{Fig4}). Traditional linear modal parameters such as fundamental frequency ($f_0$), linear damping, and spectral FWHM (Fig.~\ref{Fig4}(e-g)) fail to provide reliable biomarkers for water stress ($p > 0.05$). Similarly, visual tracking of leaf elevation angles (Fig.~\ref{Fig4}(a)) does not yield a strongly reliable early indicator ($p = 0.44$) due to highly variable turgor loss strategies among individual plants. In contrast, pigmentary changes derived from RGB imaging (Fig.~\ref{Fig4}(b-d)) and targeted SPAD values (Fig.~\ref{Fig4}(h)) shift significantly. Crucially, the nonlinear damping parameters extracted via SSM reduction, both $\beta$ and $\mathcal{K}$ (Fig.~\ref{Fig4}(j-k)), demonstrate a highly significant sign reversal ($p = 0.002$), serving as definitive, geometry-independent biomarkers for the onset of severe drought stress. The summary forest plot (Fig.~\ref{Fig4}(l)) confirms that the mechanical bifurcation isolated via SSM reduction unmasks the plant's biological drought resilience, outperforming traditional structural metrics and providing a rigorous, non-destructive phenotyping tool.

\section{Conclusion and Future Work}

We have demonstrated a rapid, remote, and non-destructive method for detecting early drought stress using the nonlinear vibration signatures of soybean stems. By applying the spectral submanifold (SSM) reduction,  we have identified a critical transition in the system's damping behavior: a shift from fluid-dominated superlinear damping to friction-dominated sublinear damping. This mechanical sign change, represented by the nonlinear damping ratio $\mathcal{K}$ and coefficient $\beta$, directly signals the loss of turgor pressure and structural dehydration.

In evaluating potential indicators for early drought stress, we identified SPAD, colorimetric indices, and the nonlinear damping parameters ($\beta$ and $\mathcal{K}$) as the most reliable metrics. While SPAD accurately measures relative chlorophyll content, it is highly labor-intensive, often requiring 10 to 30 minutes per plant to manually scan multiple wavelengths across individual leaves. Furthermore, the position-dependent survival strategy of the plant, where young leaves maintain or increase chlorophyll while mature leaves senesce, complicates the whole-plant stress assessment using localized SPAD measurements. Colorimetric analysis provides a rapid, non-invasive alternative; however, its accuracy is heavily dependent on consistent external lighting. While our laboratory experiments utilized controlled illumination, field applications with fluctuating ambient light pose significant challenges for reliable colorimetric phenotyping. In contrast, the nonlinear dynamical parameters ($\beta$ and $\mathcal{K}$) proposed here offer a simple, non-invasive, and rapid assessment of the plant's global mechanical state. By extracting these metrics from brief video recordings of structural vibrations, the entire measurement process can be completed in under a minute, providing a highly scalable and robust quantification of water stress levels.

This study provides the first experimental evidence that SSM theory can successfully extract consistent physics-based reduced-order models from the noisy, high-dimensional video data of living organisms. By focusing on the intrinsic nonlinear backbone of the dynamics, the SSM acts as a powerful filter, isolating the plant's internal structural health from environmental perturbations. Because the dimensionless nonlinear damping ratio ($\mathcal{K}$) is independent of absolute frequency shifts, localized leaf variations, or lighting conditions, this method is particularly robust for field applications where plant geometry and environment vary significantly.

Although our study focused on soybean (\textit{Glycine max}), the fundamental mechanics of turgor loss are conserved in vascular plants. The transition from viscoelastic to frictional damping promises to be a universal feature of drying plant stems. Future work will focus on expanding this SSM-based methodology to other stemmed crops, such as maize, wheat, or sorghum, and testing the robustness of $\mathcal{K}$ as a predictive tool for plant recovery following rehydration. Additionally, the rapid and non-destructive nature of camera-based data acquisition makes this method scalable. Future implementations could utilize field-mounted cameras to simultaneously observe the vibrational responses of a large number of crops without physically interfering with the plants. In field settings, plants are continually perturbed by wind, rain, neighboring plants, and handling during routine cultivation. Rather than treating these motions as nuisance variability, future systems could exploit them as natural excitation events: recording short post-perturbation decay windows, reconstructing the dominant SSM dynamics, and track hydration-sensitive normal-form coefficients over time. This would remove the need for plant-by-plant laboratory stimulation and could enable passive or minimally actuated monitoring of large crop populations. More broadly, the same framework could be extended beyond soybean stems to leaves, fruit-bearing branches, and other mechanically responsive organs, with the specific SSM coefficients serving as species-dependent reporters of internal state. In this view, the contribution is not merely a new drought metric for soybean, but a general strategy for non-destructive dynamical phenotyping of living structures from video-recorded relaxation dynamics.
Ultimately, the data-driven SSM-reduction approach used here is expected to open new avenues for smart agriculture, enabling the development of robotic phenotyping systems that can physically probe plants to assess their hydration state in real-time.

\section{Methods}

\subsection{Biological Experiments}
\textit{{Plant Material and Growth Conditions}}: 
Soybean (\textit{Glycine max} L.) seeds were purchased from Mountain Valley Seed Company (Salt Lake City, Utah, USA). Seeds were planted in an all-purpose potting soil mix (Miracle-Gro, Marysville, Ohio, USA) and grown inside a VIVOSUN grow tent (Ontario, California, USA). Plants were cultivated under a 12-hour light / 12-hour dark photoperiod using an LED grow light (VIVOSUN, Ontario, California, USA). Soil moisture was maintained by watering daily or every other day, as needed, to ensure the soil remained moist but not waterlogged. Growth conditions were maintained at room temperature ($20-25^{\circ}\text{C}$ or specified range) with a relative humidity ($\text{RH}$) controlled between $40\%$ and $60\%$.
\\

\textit{Dynamic Vibration Testing under Healthy and Drought Conditions}: 
Soybean samples were subjected to mechanical vibration testing across developmental stages, ranging from VC (First unifoliolate leaf) to R1 (Beginning bloom). To induce mechanical vibration, the plant was manually displaced in two ways; The top of the plant was gently pushed and pulled and the middle part of the main stem was gently pushed to the side. Each displacement motion was applied four to five times. Plant movement was recorded using a Photron FASTCAM NOVA S high-speed camera (Photron, Tokyo, Japan) equipped with a 50mm Nikkor lens (Nikon Corporation, Tokyo, Japan) at a capture rate of 250 frames per second (fps). Then, vibration analysis was performed using Tracker software. 

Each plant served as its own control and was first tested under healthy (fully-watered) conditions. Subsequently, drought stress was induced by withholding watering for an initial period of two or three days. The vibration test was then repeated as the plant experienced progressive dehydration. Experiments continued until the plants reached a state of irreversible wilting (defined by permanently falling and/or curled leaves).
\\

\textit{Physiological Measurements}: 
Physiological changes, concurrent with moisture loss, were monitored using a MultispeQ handheld device (PhotosynQ, East Lansing, Michigan, USA). Measurements were conducted in accordance with the manufacturer's general operating procedures. For each measurement, the device was gently clipped onto three distinct sites on the left side of a fully expanded leaf. Data collection commenced immediately upon clipping and continued until the MultispeQ device signaled completion.

\subsection{SSM Analysis}

% Motivation for nonlinear analysis + SSMs

Perturbations of the main stem provide a nonintrusive means to probe each plant's response throughout the dehydration experiment. The evolution of planar stem tip displacements across days shows a dependence of the amplitude on the frequency. This observation is incompatible with linear dynamics and hence reveals a need for nonlinear reduced order modeling. \\

To this end, we seek a reduced model that captures the nonlinear decay observed in the dominant dynamics of these systems. In the terminology of spectral submanifolds (SSMs) \cite{Haller2016}, this can be achieved by reducing the full dynamics to  the unique smoothest invariant manifold of the full nonlinear system that is tangent to the dominant spectral subspace at the unforced equilibrium of the plant. This subspace is spanned by the slowest decaying solutions of the full system linearized at the equilibrium.  The remaining faster-decaying linear modes  align exponentially fast with the trajectories on the SSM.  The existence, uniqueness, and smoothness of SSMs under  hyperbolicity and nonresonance conditions are reviewed in the SI. More details on SSM reduction and its various applications to  equations and data can be found in \cite{haller2025SSM}.

A spectral subspace $E$ associated with the slowest decaying eigenvalues is the ideal candidate for model reduction. Specifically, for an unforced nonlinear oscillatory system with $k$ dominant linear modes, a $2k$-dimensional SSM, $W(E)$, captures the observed oscillatory decay, while the remaining modes align exponentially fast with the motion on the manifold. When the eigenvalues within $E$  satisfy  low-order nonresonance conditions with respect to the eigenvalues of the faster modes outside $E$, the corresponding  slow SSM $W(E)$ is the unique smoothest, attracting, $2k$-dimensional invariant manifold of the full nonlinear system that is tangent to $E$ at the equilibrium. 

In the context of plant oscillations, trajectories initiated from different stem deflections approach the slow SSM,  $W(E)$, rapidly along the more strongly damped transverse directions. Therefore, the internal dynamics of $W(E)$ capture the amplitude-dependent decay and frequency modulation that dominate the long term response. We restrict our analysis to the portion of $W(E)$ that  can be represented as a graph over the spectral subspace $E$. Any trajectory initiated by a finite perturbation converges rapidly to  $W(E)$  and subsequently evolves along it, with its decay and frequency fully determined by the reduced dynamics on  $W(E)$ . \\

Each plant  is treated as a parametrized dynamical system, with the measurement day acting as a parameter. The system thus has a day-dependent equilibrium, linearized spectrum, and nonlinearities.  This modeling view reflects the  ongoing growth of juvenile plants and their progressive structural changes under water stress. The hyperbolicity condition of SSM theory will hold with probability one for any fixed point of a generic plant. The nonresonance condition on the linearized spectrum also holds if there is no observed modal interaction, i.e., only one mode prevails in decaying plant oscillations after the initial transients. Under these conditions, there exists a unique smoothest 2D SSM tangent to the slowest modal subspace at the plant equilibrium. The 2D ODE describing the internal dynamics of this SSM provides the simplest nonlinear description of the observed dynamics. \\

% Delay embedding
Our measurements consist of planar displacements of the stem tip $s (t) = (x(t),y(t))$ extracted from videos. For each plant-measurement day configuration, however, the true dynamics evolve in a higher-dimensional phase space that is not directly accessible. Since available planar displacement is insufficient to reconstruct the invariant manifolds for the underlying plant system, we invoke Takens' delay-embedding theorem and its extensions \cite{Takens1981, Deyle2011}. These results ensure that, for a generic observable,  a $d$-dimensional invariant set is smoothly embedded into the space of subsequent $p$ delayed observations as long as $p\geq 2d+1$.
In particular, we form delay vectors $\Phi_p(t) = \Big(s(t),\ s(t+\tau), ...,  s(t+(p-1)\tau)\Big)$ from uniformly sampled observations which provide a faithful, locally diffeomorphic representation of the dynamics in the observable space. For a 2D SSM ($d=2$) relevant to the present plant systems, the minimal embedding dimension $p=5$ is generically sufficient to obtain the SSM as a smooth manifold in the delay embedding space. In this space, the slow SSM appears as a smooth 2D invariant surface attracting all transient trajectories. The sampling intervals $\tau$ used to construct the delays for each configuration are reported in the SI. \\

%SSM Learn -> ROM
The data-driven SSMLearn algorithm \cite{cenedese2022data} identifies the 2D slow SSM in the 5D delay embedding space via regression to the part of the trajectory data whose Fourier analysis yields only one dominant frequency. SSMLearn also identifies the SSM-reduced dynamics by regressing a 2D polynomial ODE to the projection of nearby evolving trajectories on the underlying slow spectral subspace. As the final step in deriving an SSM-reduced model, SSMLearn brings the polynomial SSM dynamics into an extended normal form, i.e., constructs a smooth and invertible coordinate change that eliminates as many terms as possible from the SSM-reduced polynomial ODE model.  When expressed in polar coordinates, the resulting SSM-reduced normal form provides a compact, physically interpretable description of the plant’s effective stiffness and damping at a given hydration state, without any explicit knowledge of the full governing equations.

For an oscillatory SSM of dimension $d=2m$, the extended normal form takes the general form
\begin{equation*}
\begin{split}
        \dot \rho &= \alpha_j ( \rho, \theta) \rho_j, \\
    \dot \theta &= \omega_j (\rho, \theta), \quad j = 1, \dots m, 
\end{split}
\end{equation*}
with amplitude $\rho \in \mathbb{R}^m_+$ and angle $\theta \in \mathbb T^m $. When the linearized frequencies satisfy the appropriate nonresonance conditions, the functions $\alpha_j$ and $ \omega_j$ only depend on $\rho$. This is the case for all plant–day systems considered here.
In our specific setting, we select $m=1$ and restrict the polynomial order of the extended normal form to four in $\rho$. This yields the more specific normal form
\begin{equation}
\begin{split}
    \dot \rho &= \alpha \rho + \beta \rho^3, \\
    \dot \theta & = \theta_0 + \gamma \rho^2,
\end{split}
\end{equation}
which formally agrees the normal form obtained for Hopf bifurcations on center manifolds (also known as the Stuart--Landau equations in fluid mechanics). Higher-order truncations of the SSM-reduced normal form generally improve accuracy but quickly lead to overfit, given that a low number of trajectories are realistically available on each day.  With the chosen cibic order, unseen test trajectories, obtained from initial stem deflections yield normalized mean-trajectory errors in the range of $1\%-5]\%$ across all plant–day configurations. These low errors indicate that the leading nonlinear contributions are accurately captured across all plant–day configurations. More detailed results are given in the SI.

The coefficients of $\alpha(\rho)$ summarize the effective damping properties of each plant at the corresponding hydration state. Across individual plants and days under water deprivation, we observe a systematic evolution of the nonlinear damping coefficients, including a sign change of the leading nonlinear damping coefficient $\beta$. This evolution of the cubic damping term reflects a hydration-dependent modification of the effective stiffness of the plant. It signals a dissipative response that cannot be accounted for by changes in the linear decay rate alone.

% ---------------------------------------------------------------------
\subsection{Statistical Analysis}
% ---------------------------------------------------------------------

\textit{Experimental design and hydration stages:} Every plant served as its own longitudinal control, so all comparisons used a paired design. Days are reported relative to the cessation of watering ($\mathrm{Day}=0$). Each plant--day observation was assigned to one of three hydration stages: healthy ($\mathrm{Day}\le 0$), {intermediate} ($1\le \mathrm{Day}\le 5$), and {stressed} ($\mathrm{Day}\ge 6$). The same stage boundaries were applied to every morphological, pigmentary, spectral, and SSM-derived metric.

\textit{Per-plant aggregation:} For each metric, plant, and hydration stage, we first averaged all observations falling within that stage (pooling over the repeated measurement days, and over leaves where a metric was leaf-resolved) to obtain a single representative value per plant per stage. This aggregation removes the pseudoreplication that would otherwise arise from treating multiple leaves or repeated days of the same plant as independent samples, and it makes the unit of statistical analysis the individual plant ($n=10$ plants).

\textit{Significance test:} Healthy and stressed states were compared with the two-sided Wilcoxon signed-rank test, a non-parametric paired test applied to the per-plant stage means and paired plant-by-plant. The non-parametric test was chosen in preference to a paired $t$-test because the per-plant sample is small and the metrics are not assumed to be normally distributed; the signed-rank test is robust to outliers and to the heavy-tailed distributions observed for several mechanical metrics. Only plants measured in {both} the healthy and the stressed stage were included in a given test, and comparisons supported by fewer than four paired plants were reported as not available rather than as a $p$-value. A two-sided significance threshold of $\alpha=0.05$ was used throughout, and $p$-values are reported to two significant figures. Where indicated, effect sizes are given as Cohen's~$d$ computed from the pooled standard deviation of the two stages. The summary forest plot (Fig.~\ref{Fig4}(l)) ranks all metrics by their healthy-versus-stressed signed-rank $p$-value.

\textit{Implementation:} All statistical computations were performed in Python~3 using \texttt{scipy.stats.wilcoxon} and \texttt{numpy}; identical code paths were used for the morphological (Fig.~\ref{Fig1}), kinematic/spectral (Fig.~\ref{Fig2}), SSM (Fig.~\ref{Fig3}), and summary (Fig.~\ref{Fig4}) panels so that every reported $p$-value reflects the same test, the same stage definitions, and the same per-plant aggregation.

% ---------------------------------------------------------------------
\subsection{Video Tracking and Signal Conditioning}
% ---------------------------------------------------------------------

\textit{Kinematic tracking:} Free-vibration sequences recorded at $250\,\mathrm{fps}$ were analyzed by tracking fiducial points on the main stem, the stem apex, and the first three leaf tip/base pairs. Points were initialized on the first frame and propagated through the sequence with a pyramidal Lucas--Kanade optical-flow tracker; image contrast was enhanced with contrast-limited adaptive histogram equalization (CLAHE) prior to tracking. Pixel trajectories were converted to physical units using a per-image calibration ($\mathrm{cm}\,\mathrm{px}^{-1}$) obtained from a scale bar of known length in each recording.

\textit{Signal conditioning:} For each tracked point we formed the planar displacement $s(t)=(x(t),y(t))$ relative to the stem base. The equilibrium (rest) position was estimated as the mean of the final portion of each decay trace and subtracted, so that the analyzed signal oscillates about the origin. 

% ---------------------------------------------------------------------
\subsection{Spectral and Linear-Damping Feature Extraction}
% ---------------------------------------------------------------------

\textit{Fundamental frequency:} The power spectrum of each decay trace was obtained from a one-sided fast Fourier transform (Fig.~\ref{Fig2}(h)), and the fundamental frequency $f_0$ was taken as the location of the dominant peak within a physiologically plausible band ($0.5$--$10\,\mathrm{Hz}$). Short-time Fourier transforms (Fig.~\ref{Fig2}(i,j)) were used to visualize the transient frequency content during the decay.

\textit{Linear damping rate:} The linear decay rate was estimated by the logarithmic-decrement method: successive oscillation-peak amplitudes were detected, and the natural logarithm of the peak magnitude was regressed against {time} (in seconds, i.e.\ peak index divided by the sampling rate). The negative slope of this fit yields the decay rate $\delta=\zeta\omega_n$ in units of $\mathrm{s}^{-1}$ (Fig.~\ref{Fig2}(k),
Fig.~\ref{Fig4}(f)); reporting against time rather than sample index makes the estimate independent of frame rate.

\subsection{Delay Embedding and Nonlinear Damping Ratio}

\textit{Delay coordinates:} For visualization of the reconstructed attractors (Fig.~\ref{Fig3}(a--c)) we formed three-dimensional delay vectors $\big(s(t),\,s(t+T),\,s(t+2T)\big)$ from the equilibrium-subtracted stem displacement. The delay was set to a quarter of the dominant oscillation period, $T=T_0/4=1/(4f_0)$, which corresponds to a $90^{\circ}$ phase shift and therefore unfolds a near-sinusoidal orbit into a maximally opened (approximately circular) trajectory; equivalently, $T_0/4$ is the first zero crossing of the autocorrelation of a single-frequency signal. %(The five-dimensional delay embedding used for the SSM identification, and the per-configuration sampling intervals, are described in Section~IV.2 and the SI.)

\textit{Nonlinear damping ratio:} From the normal-form coefficients $\alpha$ (linear) and $\beta$ (cubic) of each plant--day model [Eq.~(\ref{eq:decay})], we defined the dimensionless nonlinear damping ratio $\mathcal{K}=(\beta/\alpha)\,L^{2}$, where $L$ is the soil-to-leaf-base  distance for that plant and day; the factor $L^{2}$ non-dimensionalizes the ratio and removes its dependence on absolute plant size, enabling comparison across developmental stages.

\subsection{Colorimetric and Morphological Quantification}

\textit{Colorimetry:} Leaf color was quantified from segmented leaf tissue as the mean per-channel intensity and the normalized chromaticity $r,g,b = R,G,B/(R+G+B)$ (Fig.~\ref{Fig1}(h)). For the stage comparisons in
Fig.~\ref{Fig4}(b--d) we report the hue angle, the greenness index $(G-R)/(G+R)$, and the normalized green fraction $G/(R+G+B)$.

\textit{Morphology:} Lamina angle (relative to the vertical), lamina length, and petiole length (Fig.~\ref{Fig1}(e--g)) were measured from the calibrated images; values are summarized as mean~$\pm$~standard error of the mean across plants unless otherwise noted.

\section{{Data avaiability:}} Source data are provided with this paper. Additionally, all data and Matlab scripts are available at \\
(DOI:10.17605/OSF.IO/4RS9W).

\section{Acknowledgements}
The authors thank Ashok Kumar and Arijit Saha for plant-recovery experiments.

\section{Author contributions statement} 
S.J. and G.H. conceived the idea. J.Y. performed all biological experiments. J.Y. and I.T. prepared main figures. I.T. conducted the data and SSM analysis. S.J. and I.T. wrote the manuscript draft. All authors reviewed and revised the manuscript.

\section{Competing interests statement} The authors declare that they have no competing interests.

\bibliographystyle{ieeetr}
\bibliography{apssamp}

@book{niklas1992plant,
  title={Plant biomechanics: an engineering approach to plant form and function},
  author={Niklas, Karl J},
  year={1992},
  publisher={University of Chicago press}
}

@book{haller2025SSM,
  title={Modeling Nonlinear Dynamics from Equations and Data - with applications to Solids, Fluids, and Controls},
  author={Haller, G.},
  year={2025},
  publisher={SIAM Presse, Philadelphia, PA}
}

@article{kanahama2023rigidity,
  title={Rigidity control mechanism by turgor pressure in plants},
  author={Kanahama, Tohya and Tsugawa, Satoru and Sato, Motohiro},
  journal={Scientific Reports},
  volume={13},
  number={1},
  pages={2063},
  year={2023},
  publisher={Nature Publishing Group UK London},
  doi={10.1038/s41598-023-29294-5},
  url={https://doi.org/10.1038/s41598-023-29294-5}
}

@article{manavalan2009physiological,
  title={Physiological and molecular approaches to improve drought resistance in soybean},
  author={Manavalan, Lakshmi P and Guttikonda, Satish K and Tran, Lam-Son Phan and Nguyen, Henry T},
  journal={Plant and Cell Physiology},
  volume={50},
  number={7},
  pages={1260--1276},
  year={2009},
  publisher={Oxford University Press}
}

@article{franco2025advancements,
  author  = {Franco-Navarro, Juan D. and Padilla, Yaiza Gara and {\'A}lvarez, Sara and Calatayud, {\'A}ngeles and Colmenero-Flores, Jos{\'e} Manuel and G{\'o}mez-Bellot, Mar{\'i}a Jos{\'e} and Hern{\'a}ndez, Jos{\'e} Antonio and Mart{\'i}nez-Alcal{\'a}, Isabel and Penella, Consuelo and P{\'e}rez-P{\'e}rez, Juan Gabriel and S{\'a}nchez-Blanco, Mar{\'i}a Jes{\'u}s and Tasa, Mar{\'i}a and Acosta-Motos, Jos{\'e} Ram{\'o}n},
  title   = {Advancements in Water-Saving Strategies and Crop Adaptation to Drought: A Comprehensive Review},
  journal = {Physiologia Plantarum},
  year    = {2025},
  volume  = {177},
  number  = {4},
  pages   = {e70332},
  doi     = {10.1111/ppl.70332},
  url     = {https://doi.org/10.1111/ppl.70332},
  publisher = {John Wiley & Sons Ltd}
}

@article{biswas2025water,
  author  = {Biswas, Auindrila and Sarkar, Sourakanti and Das, Sumanta and Dutta, Suman and Choudhury, Malini Roy and Giri, Anmol and Bera, Bimal and Bag, Koushik and Mukherjee, Bishal and Banerjee, Koushik and Gupta, Debaditya and Paul, Debasish},
  title   = {Water scarcity: A global hindrance to sustainable development and agricultural production -- A critical review of the impacts and adaptation strategies},
  journal = {Cambridge Prisms: Water},
  year    = {2025},
  volume  = {3},
  pages   = {e4},
  doi     = {10.1017/wat.2024.16},
  url     = {https://doi.org/10.1017/wat.2024.16},
  publisher = {Cambridge University Press}
}

@article{scholander1965sap,
  title={Sap pressure in vascular plants},
  author={Scholander, Per Fredrik and Hammel, HT and Bradstreet, ED and Hemmingsen, EA},
  journal={Science},
  volume={148},
  number={3668},
  pages={339--346},
  year={1965},
  publisher={American Association for the Advancement of Science},
  doi={10.1126/science.148.3668.339}
}

@article{brodribb2017evolution,
  title={Evolution of the stomatal regulation of plant water content},
  author={Brodribb, Timothy J and McAdam, Scott AM},
  journal={Plant Physiology},
  volume={174},
  number={2},
  pages={639--649},
  year={2017},
  publisher={American Society of Plant Biologists},
  doi={10.1104/pp.17.00578}
}

@incollection{jones2004application,
  title={Application of thermal imaging and infrared sensing in plant physiology and agriculture},
series = {Advances in Botanical Research},
publisher = {Academic Press},
volume = {41},
pages = {107-163},
year = {2004},
booktitle = {Incorporating Advances in Plant Pathology},
issn = {0065-2296},
doi = {https://doi.org/10.1016/S0065-2296(04)41003-9},
author = {Hamlyn G. Jones}
}

@article{burgess2001improved,
  title={An improved heat pulse method to measure low and reverse rates of sap flow in woody plants},
  author={Burgess, Stephen SO and Adams, Mark A and Turner, Neil C and Beverly, Christopher R and Ong, Chin K and Khan, AA and Bleby, TM},
  journal={Tree Physiology},
  volume={21},
  number={9},
  pages={589--598},
  year={2001},
  publisher={Oxford University Press},
  doi={10.1093/treephys/21.9.589}
}

@incollection{GAONKAR201973,
title = {Chapter 4 - Soybean},
editor = {Zhongli Pan and Ruihong Zhang and Steven Zicari},
booktitle = {Integrated Processing Technologies for Food and Agricultural By-Products},
publisher = {Academic Press},
pages = {73-104},
year = {2019},
isbn = {978-0-12-814138-0},
doi = {https://doi.org/10.1016/B978-0-12-814138-0.00004-6},
url = {https://www.sciencedirect.com/science/article/pii/B9780128141380000046},
author = {Varsha Gaonkar and Kurt A. Rosentrater} 
}

@incollection{pagano2016importance,
  title={The Importance of Soybean Production Worldwide},
  author={Pagano, M. R. and Miransari, M.},
  booktitle={Abiotic and Biotic Stresses in Soybean Production},
  pages={1--26},
  year={2016},
  publisher={Academic Press},
  doi={10.1016/B978-0-12-801535-3.00001-X}
}

@article{gart2015droplet,
  title={Droplet impacting a cantilever: A leaf-raindrop system},
  author={Gart, Sean and Mates, Joseph E and Megaridis, Constantine M and Jung, Sunghwan},
  journal={Physical Review Applied},
  volume={3},
  number={4},
  pages={044019},
  year={2015},
  publisher={APS},
  doi={10.1103/PhysRevApplied.3.044019}
}

@article{py2006frequency,
  title={A frequency lock-in mechanism in the interaction between wind and crop canopies},
  author={Py, Charlotte and de Langre, Emmanuel and Moulia, Bruno},
  journal={Journal of Fluid Mechanics},
  volume={568},
  pages={425--449},
  year={2006},
  publisher={Cambridge University Press},
  doi={10.1017/S0022112006002667}
}

@article{delangre2008effects,
  title={Effects of wind on plants},
  author={de Langre, Emmanuel},
  journal={Annual Review of Fluid Mechanics},
  volume={40},
  pages={141--168},
  year={2008},
  publisher={Annual Reviews},
  doi={10.1146/annurev.fluid.40.111406.102135}
}

@book{oecdfao2023outlook,
  title={OECD-FAO Agricultural Outlook 2023-2032},
  author={{OECD/FAO}},
  year={2023},
  publisher={OECD Publishing, Paris},
  doi={10.1787/d00257bc-en},
  url={https://doi.org/10.1787/d00257bc-en}
}

@article{unchronicle2023water,
  title={Water Scarcity, the Climate Crisis and Global Food Security: A Call for Collaborative Action},
  author={{United Nations}},
  journal={UN Chronicle},
  year={2023},
  url={https://www.un.org/en/un-chronicle/water-scarcity-climate-crisis-and-global-food-security-call-collaborative-action},
  note={Accessed: 2024-05-22}
}

@article{graham2003legumes,
  title={Legumes: Importance and Constraints to Greater Use},
  author={Graham, P. H. and Vance, C. P.},
  journal={Plant Physiology},
  volume={131},
  number={3},
  pages={872--877},
  year={2003},
  doi={10.1104/pp.102.017772}
}

@article{mutava2015understanding,
  author  = {Mutava, Raymond N. and Prince, Silvas Jebakumar K. and Syed, Naeem Hasan and Song, Li and Valliyodan, Babu and Chen, Wei and Nguyen, Henry T.},
  title   = {Understanding abiotic stress tolerance mechanisms in soybean: A comparative evaluation of soybean response to drought and flooding stress},
  journal = {Plant Physiology and Biochemistry},
  year    = {2015},
  volume  = {86},
  pages   = {109--120},
  doi     = {10.1016/j.plaphy.2014.11.010},
  url     = {https://doi.org/10.1016/j.plaphy.2014.11.010},
  publisher = {Elsevier}
}

@article{kanahama2023mechanics,
  title={Mechanics-based classification rule for plants},
  author={Kanahama, Tohya and Sato, Motohiro},
  journal={Proceedings of the National Academy of Sciences},
  volume={120},
  number={41},
  pages={e2308319120},
  year={2023},
  publisher={National Acad Sciences},
  doi={10.1073/pnas.2308319120},
  url={https://doi.org/10.1073/pnas.2308319120}
}

@article{Chenetal2015,
  author  = {Chen, Daoqian and Wang, Shiwen and Xiong, Binglin and Cao, Beibei and Deng, Xiping},
  title   = {Carbon/Nitrogen Imbalance Associated with Drought-Induced Leaf Senescence in Sorghum bicolor},
  journal = {PLOS ONE},
  year    = {2015},
  volume  = {10},
  pages   = {e0137026},
  doi     = {10.1371/journal.pone.0137026}
}

@article{Tanetal2023,
  author  = {Tan, Shuya and Sha, Yueqi and Sun, Liwei and Li, Zhonghai},
  title   = {Abiotic Stress-Induced Leaf Senescence: Regulatory Mechanisms and Application},
  journal = {International Journal of Molecular Sciences},
  year    = {2023},
  volume  = {24},
  pages   = {11996},
  doi     = {10.3390/ijms241511996}
}

@article{dominguez2011overview,
  title={An overview on plant cuticle biomechanics},
  author={Dom{\'\i}nguez, Eva and Cuartero, Jes{\'u}s and Heredia, Antonio},
  journal={Plant Science},
  volume={181},
  number={2},
  pages={77--84},
  year={2011},
  publisher={Elsevier}
}

@article{salmen1984viscoelastic,
  title={Viscoelastic properties of in situ lignin under water-saturated conditions},
  author={Salm{\'e}n, Lennart},
  journal={Journal of Materials science},
  volume={19},
  number={9},
  pages={3090--3096},
  year={1984},
  publisher={Springer}
}

@article{sathiya2023smart,
  title={Smart agriculture: Emerging and future farming technologies},
  author={Sathiya, S and Antony, Cecil and Ghodke, Praveen Kumar},
  journal={Recent Trends and Best Practices in Industry 4.0},
  pages={135--181},
  year={2023},
  publisher={River Publishers}
}

@article{zhao2019crop,
  title={Crop phenomics: current status and perspectives},
  author={Zhao, Chunjiang and Zhang, Ying and Du, Jianjun and Guo, Xinyu and Wen, Weiliang and Gu, Shenghao and Wang, Jinglu and Fan, Jiangchuan},
  journal={Frontiers in Plant Science},
  volume={10},
  pages={714},
  year={2019},
  publisher={Frontiers Media SA}
}

@article{Dong2019,
  author    = {Dong, Shoukun and Jiang, Yingze and Dong, Yuchen and Wang, Libin and Wang, Wenjia and Ma, Zezhong and Yan, Chao and Ma, Chunmei and Liu, Lijun},
  title     = {A study on soybean responses to drought stress and rehydration},
  journal   = {Saudi Journal of Biological Sciences},
  volume    = {26},
  number    = {7},
  pages     = {2006--2017},
  year      = {2019},
  doi       = {10.1016/j.sjbs.2019.08.005}
}

@article{Haller2016,
  title={Nonlinear normal modes and spectral submanifolds: existence, uniqueness and use in model reduction},
  author={Haller, George and Ponsioen, Stefan},
  journal={Nonlinear Dynamics},
  volume={86},
  pages={1493--1534},
  year={2016},
  publisher={Springer}
}

@article{Hasanuzzaman2022,
  author    = {Hasanuzzaman, Md. and Shabala, Lana and Brodribb, Timothy J. and Zhou, Meixue and Shabala, Sergey},
  title     = {Understanding the Role of Physiological and Agronomical Traits during Drought Recovery as a Determinant of Differential Drought Stress Tolerance in Barley},
  journal   = {Agronomy},
  volume    = {12},
  number    = {9},
  pages     = {2136},
  year      = {2022},
  doi       = {10.3390/agronomy12092136}
}

@article{SinghBakala2025,
  author    = {Singh-Bakala, Harmeet and Ravelombola, Francia and Adeva, Cheryl and Oliveira, Maiara and Zhang, Ru and Argenta, Jessica and Shannon, Grover and Lin, Feng},
  title     = {Harnessing photosynthetic and morpho-physiological traits for drought-resilient soybean: integrating field phenotyping and predictive approaches},
  journal   = {Frontiers in Plant Physiology},
  volume    = {3},
  pages     = {1591146},
  year      = {2025},
  doi       = {10.3389/fphgy.2025.1591146}
}

@article{zegada2013photosynthetic,
  title={Photosynthetic response of sweet sorghum to drought and re-watering at different growth stages},
  author={Zegada-Lizarazu, Walter and Monti, Andrea},
  journal={Physiologia Plantarum},
  volume={149},
  number={1},
  pages={56--66},
  year={2013},
  publisher={Wiley Online Library},
  doi={10.1111/ppl.12016}
}

@article{cenedese2022data,
  author  = {Cenedese, Mattia and Ax{\aa}s, Joar and B{\"a}uerlein, Bastian and Avila, Kerstin and Haller, George},
  title   = {Data-driven modeling and prediction of non-linearizable dynamics via spectral submanifolds},
  journal = {Nature Communications},
  year    = {2022},
  volume  = {13},
  number  = {1},
  pages   = {872},
  doi     = {10.1038/s41467-022-28518-y},
  url     = {https://doi.org/10.1038/s41467-022-28518-y},
  publisher = {Nature Publishing Group}
}

@incollection{Takens1981,
    address = {Berlin, Heidelberg},
    title = {Detecting strange attractors in turbulence},
    volume = {898},
    copyright = {http://www.springer.com/tdm},
    isbn = {978-3-540-11171-9 978-3-540-38945-3},
    url = {http://link.springer.com/10.1007/BFb0091924},
    language = {en},
    urldate = {2024-04-06},
    booktitle = {Dynamical {Systems} and {Turbulence}, {Warwick} 1980},
    publisher = {Springer Berlin Heidelberg},
    author = {Takens, Floris},
    editor = {Rand, David and Young, Lai-Sang},
    year = {1981},
    doi = {10.1007/BFb0091924},
    note = {Series Title: Lecture Notes in Mathematics},
    pages = {366--381},
}

@article{Deyle2011,
 author = {Deyle, Ethan R. and Sugihara, George},
 title = {Generalized Theorems for Nonlinear State Space Reconstruction},
 journal = {PLOS ONE},
 year = {2011},
 doi = {10.1371/journal.pone.0018295}
}

\end{document}

\title{Supplementary Information for:\\ Non-Destructive Detection of Early Drought Stress in Soybeans via Spectral Submanifold Analysis}

\author{Sunghwan Jung}
\affiliation{Biological and Environmental Engineering, Cornell University, Ithaca, NY 14853, USA}

\author{Isabella Thiel}
\affiliation{Institute for Mechanical Systems, ETH Zurich, Zurich, Switzerland}

\author{Jisoo Yuk}
\affiliation{Biological and Environmental Engineering, Cornell University, Ithaca, NY 14853, USA}

\author{George Haller}
\affiliation{Institute for Mechanical Systems, ETH Zurich, Zurich, Switzerland}

\maketitle

\section*{Supplementary Figures}

This document contains comprehensive, per-plant datasets (Pots 2 through 11) tracking the morphological, physiological, and mechanical progression of drought stress. Each figure below summarizes the complete experimental profile for an individual plant, allowing for a direct comparison between visual wilting, chlorophyll degradation, and the shifting nonlinear damping backbone.

\textbf{Morphological and Physiological Responses:}
The visual and physiological markers of drought stress exhibit expected but highly variable trends across individual plants. Normalized leaf elevation angles generally decline following the cessation of watering (Day 0), driven by macroscopic turgor loss; however, the onset and severity of this drooping are inconsistent across different pots and canopy nodes. Conversely, the physiological response is highly systematic: SPAD values reveal a distinct position-dependent degradation. While the youngest upper leaves (e.g., L1) maintain or even slightly increase their relative chlorophyll content—reflecting a "stay-green" nutrient remobilization strategy—the mature lower leaves consistently exhibit a precipitous decline in SPAD values as progressive dehydration induces early senescence.

\textbf{Limitations of Linear Vibration Dynamics:} Analysis of the raw kinematics highlights the limitations of traditional linear modal analysis for plant phenotyping. The Fast Fourier Transform (FFT) amplitude spectra fail to provide a universal or reliable biomarker for water stress. While some plants exhibit a slight lowering of the fundamental frequency or a narrowing of spectral peaks, these trends are inconsistent across the sample set. This variance is expected, as macroscopic changes in plant geometry (e.g., leaf drooping, changes in mass distribution) unpredictably alter the global stiffness and frequency, obscuring the underlying material changes.

\textbf{Robustness of the Nonlinear Damping Backbone (SSM):}
In stark contrast to the linear frequency response, the nonlinear damping term extracted via Spectral Submanifold (SSM) reduction serves as a highly robust and universal indicator of internal drought stress. Across all tested plants (Pots 2 through 11), the isolated nonlinear damping term ($\beta$) exhibits a clear morphological transition. In the healthy, well-watered state, the damping backbone consistently curves to the left (super-linear damping, negative $\beta$). As severe drought progresses, this curvature uniformly straightens and flips to the right (sub-linear damping, positive $\beta$). This distinct sign reversal physically unmasks the plant's internal transition from a pressurized, viscoelastic fluid-structure to a dry, friction-dominated cellular solid, providing a definitive, geometry-independent metric for turgor loss.

\vspace{1cm}

% --- SUMMARY FOR POT 2 ---
\begin{figure}[h]
    \centering
    \includegraphics[width=1\textwidth]{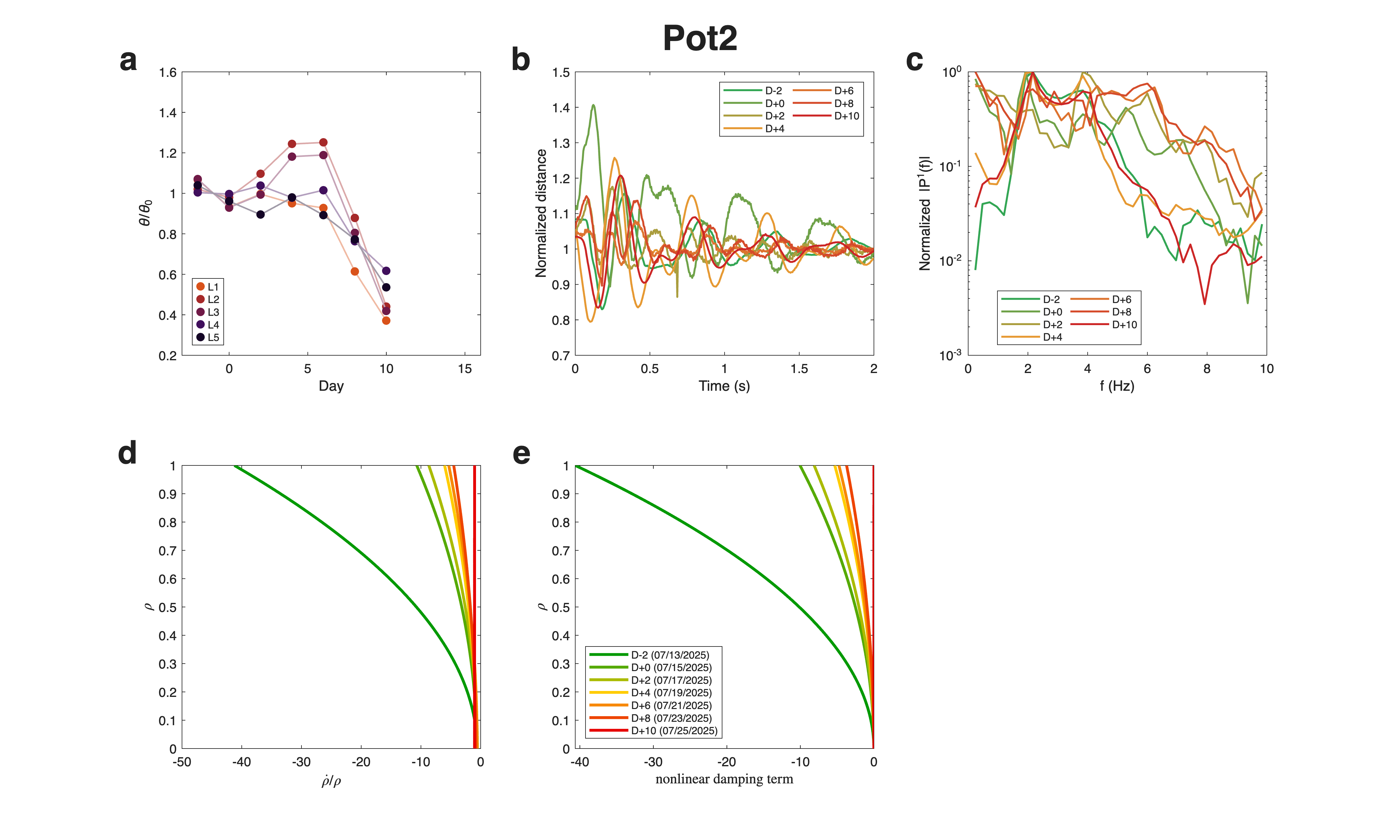}
\caption{\textbf{Comprehensive experimental summary for Pot 2.} (a) Normalized leaf elevation angles across canopy nodes over the dehydration cycle. (b) Normalized time-series relative distance between the stem base and leaf tip during free-vibration pluck tests. (c) Fast Fourier Transform (FFT) amplitude spectra of the decay responses. (d) Extracted damping backbone curves showing the total instantaneous damping rate ($\dot{\rho}/\rho$) against displacement amplitude ($\rho$). (e) The isolated nonlinear damping term. The transition from a curved (super-linear) to a straightened (linear/sub-linear) damping profile reflects the underlying loss of viscoelastic turgor.}    \label{FigS1}
\end{figure}

\clearpage

% --- SUMMARY FOR POT 3 ---
\begin{figure}[h]
    \centering
    \includegraphics[width=1\textwidth]{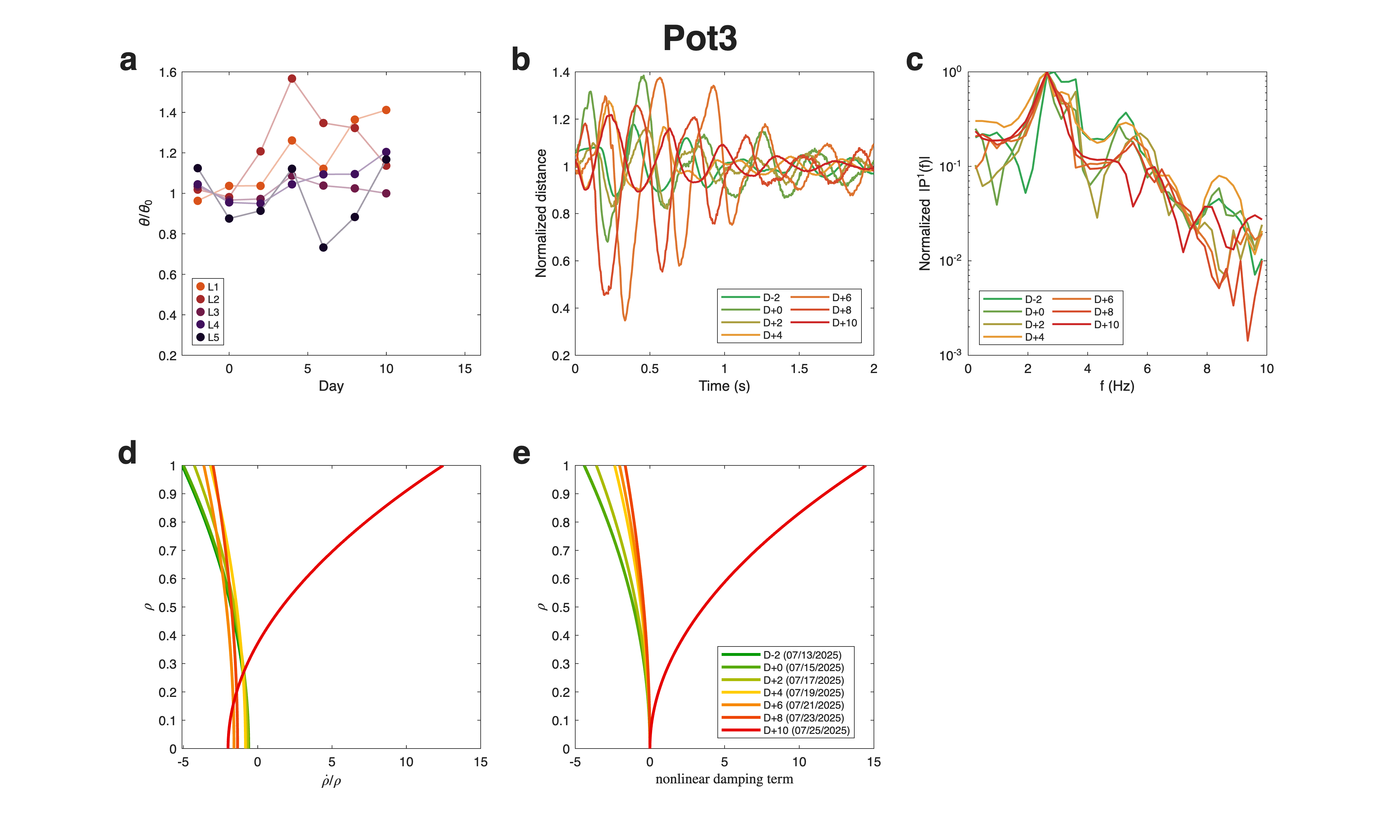}
    \caption{\textbf{Comprehensive experimental summary for Pot 3.} (a) Normalized leaf elevation angle. (b) Time-series relative displacement. (c) FFT spectrum. (d) Total instantaneous damping rate ($\dot{\rho}/\rho$). (e) Nonlinear damping term. The mechanical shift isolated by the Spectral Submanifold (SSM) reduction closely tracks the morphological onset of turgor loss.}
    \label{FigS2}
\end{figure}

\clearpage

% --- SUMMARY FOR POT 4 ---
\begin{figure}[h]
    \centering
    \includegraphics[width=1\textwidth]{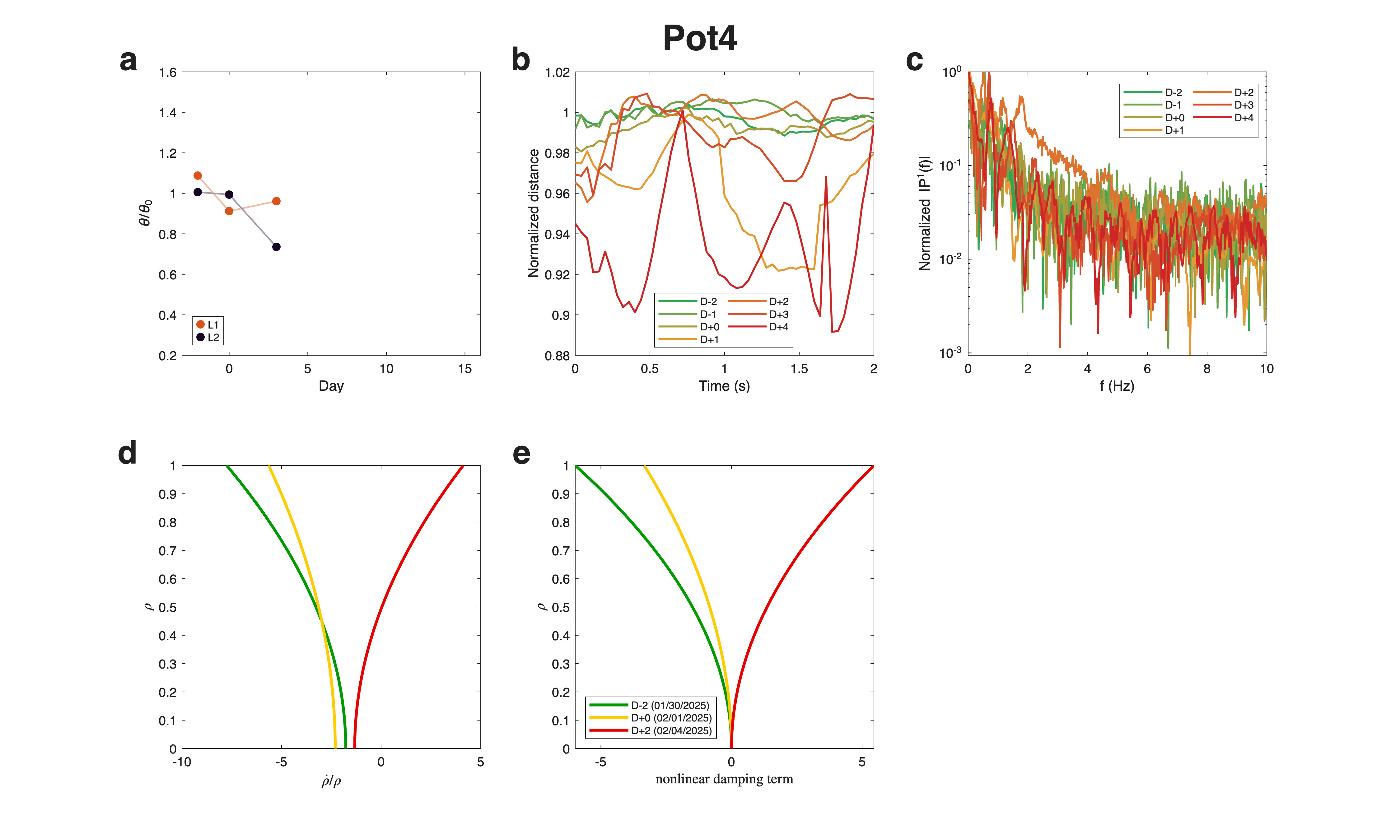}
    \caption{\textbf{Comprehensive experimental summary for Pot 4.} (a) Normalized leaf elevation angle. (b) Time-series relative displacement. (c) FFT spectrum. (d) Total instantaneous damping rate ($\dot{\rho}/\rho$). (e) Nonlinear damping term. The data demonstrate the robustness of the nonlinear damping transition across varying baseline geometries.}
    \label{FigS3}
\end{figure}

\clearpage

% --- SUMMARY FOR POT 5 ---
\begin{figure}[h]
    \centering
    \includegraphics[width=1\textwidth]{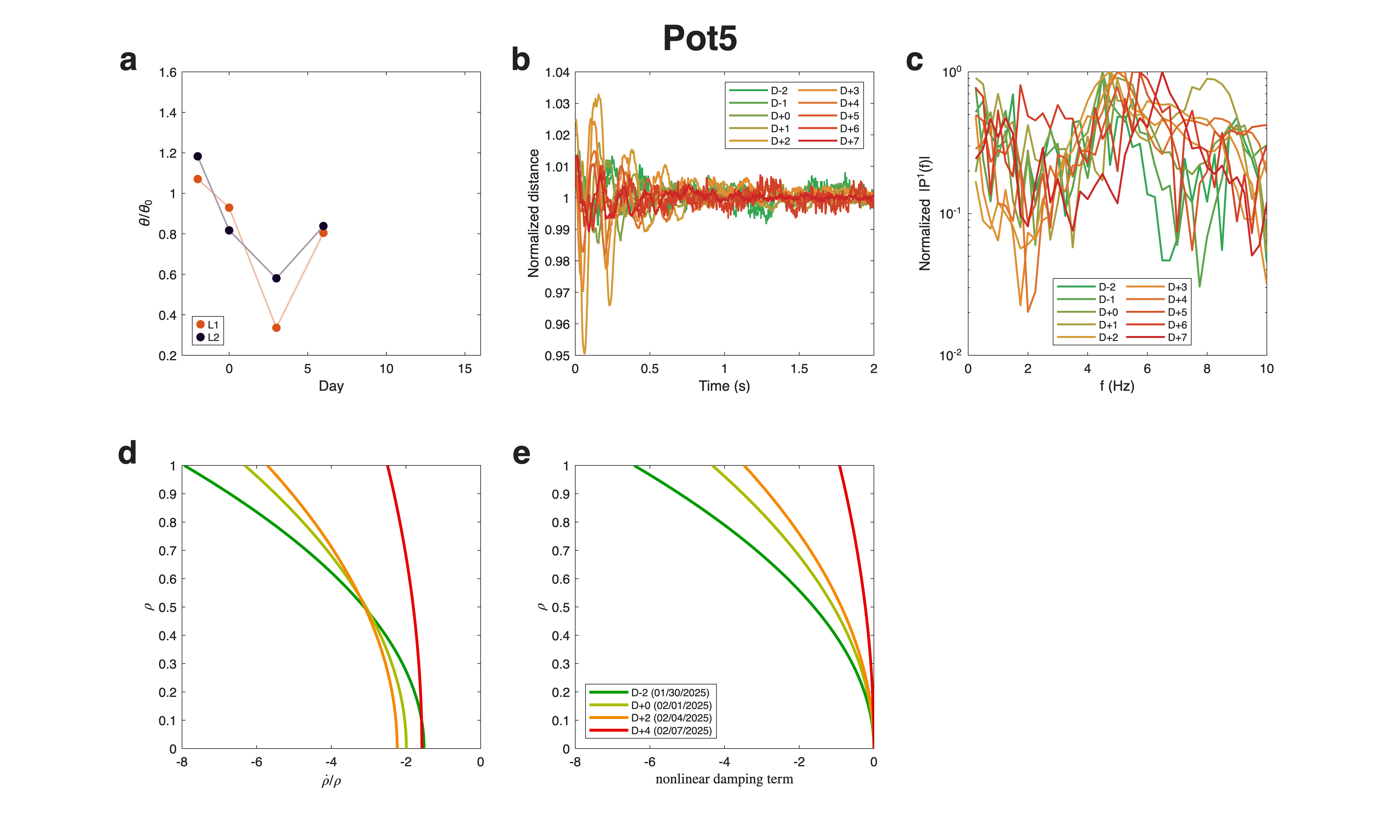}
    \caption{\textbf{Comprehensive experimental summary for Pot 5.} (a) Normalized leaf elevation angle. (b) Time-series relative displacement. (c) FFT spectrum. (d) Total instantaneous damping rate ($\dot{\rho}/\rho$). (e) Nonlinear damping term. The decay profiles and frequency responses evolve distinctly as the plant transitions to a dry, structurally stiffened state.}
    \label{FigS4}
\end{figure}

\clearpage

% --- SUMMARY FOR POT 6 ---
\begin{figure}[h]
    \centering
    \includegraphics[width=1\textwidth]{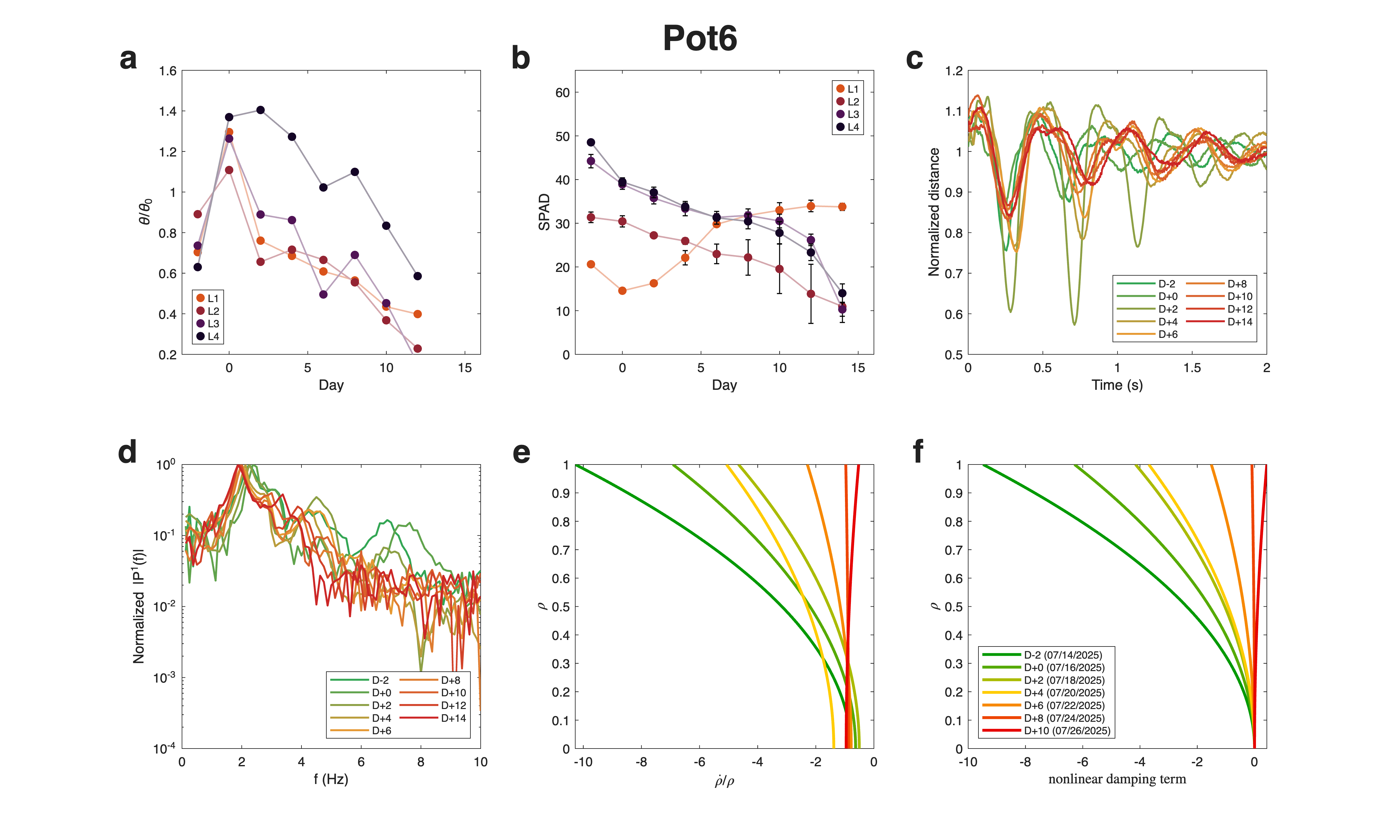}
    \caption{\textbf{Comprehensive experimental summary for Pot 6.} (a) Normalized leaf elevation angle. (b) SPAD values. (c) Time-series relative displacement. (d) FFT spectrum. (e) Total instantaneous damping rate ($\dot{\rho}/\rho$). (f) Nonlinear damping term. A consistent divergence in SPAD is observed alongside the mechanical transition from fluid viscosity to dry structural friction.}
    \label{FigS5}
\end{figure}

\clearpage

% --- SUMMARY FOR POT 7 ---
\begin{figure}[h]
    \centering
    \includegraphics[width=1\textwidth]{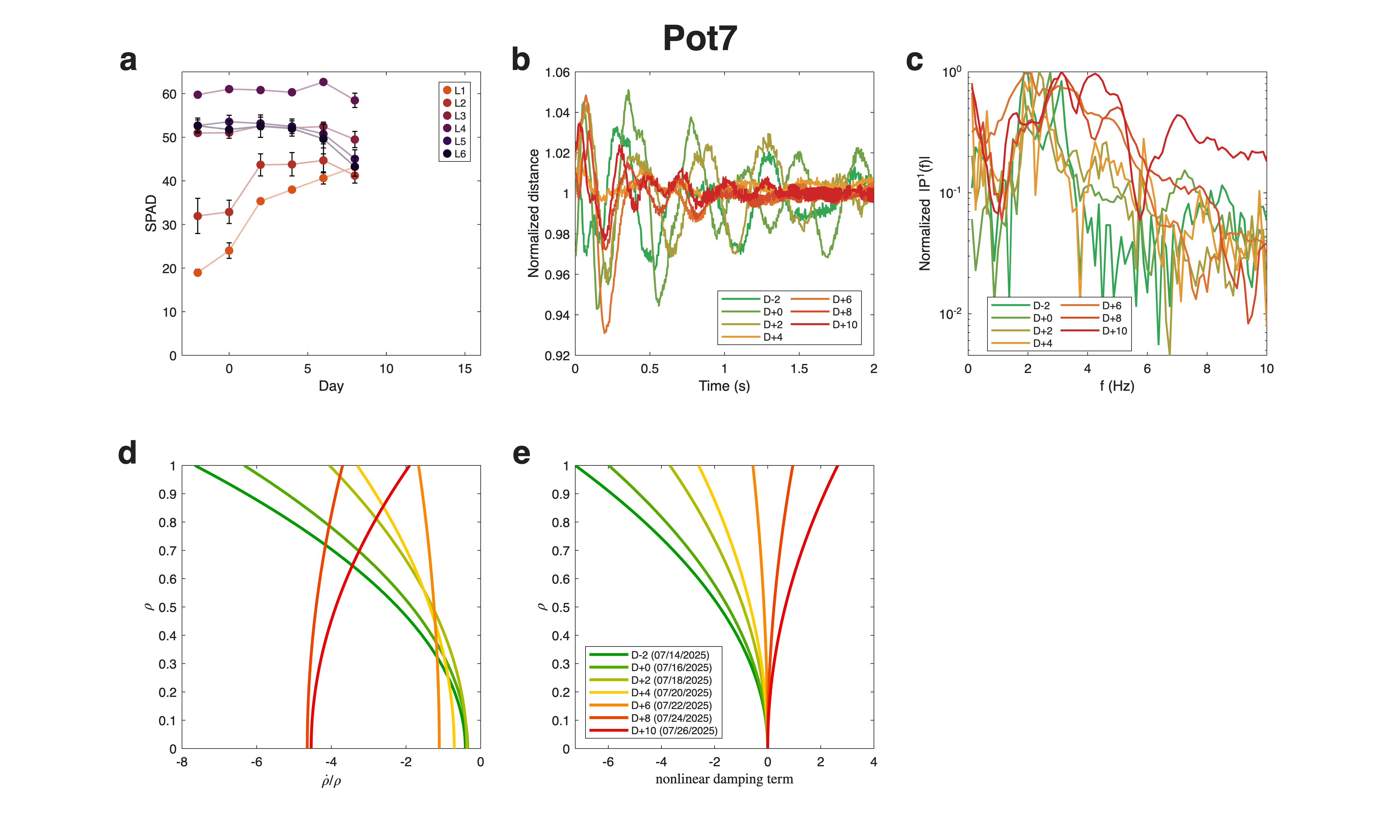}
    \caption{\textbf{Comprehensive experimental summary for Pot 7.} (a) SPAD values. (b) Time-series relative displacement. (c) FFT spectrum. (d) Total instantaneous damping rate ($\dot{\rho}/\rho$). (e) Nonlinear damping term. Amplitude-dependent damping and frequency modulation are clearly visible, confirming the necessity of a nonlinear modal analysis.}
    \label{FigS6}
\end{figure}

\clearpage

% --- SUMMARY FOR POT 8 ---
\begin{figure}[h]
    \centering
    \includegraphics[width=1\textwidth]{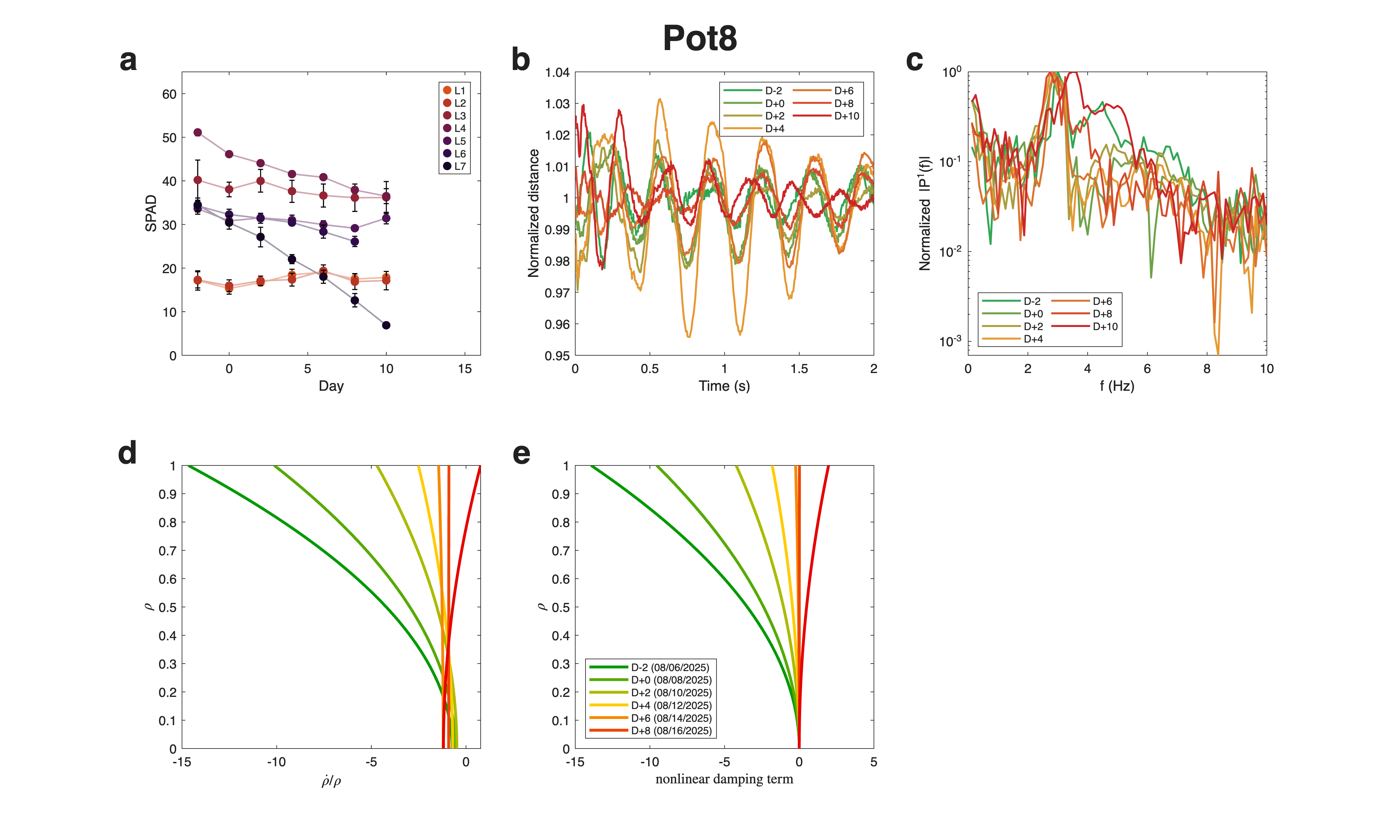}
    \caption{\textbf{Comprehensive experimental summary for Pot 8.} (a) SPAD values. (b) Time-series relative displacement. (c) FFT spectrum. (d) Total instantaneous damping rate ($\dot{\rho}/\rho$). (e) Nonlinear damping term. The progression of drought stress systematically alters both the temporal decay envelope and the damping backbone.}
    \label{FigS7}
\end{figure}

\clearpage

% --- SUMMARY FOR POT 9 ---
\begin{figure}[h]
    \centering
    \includegraphics[width=1\textwidth]{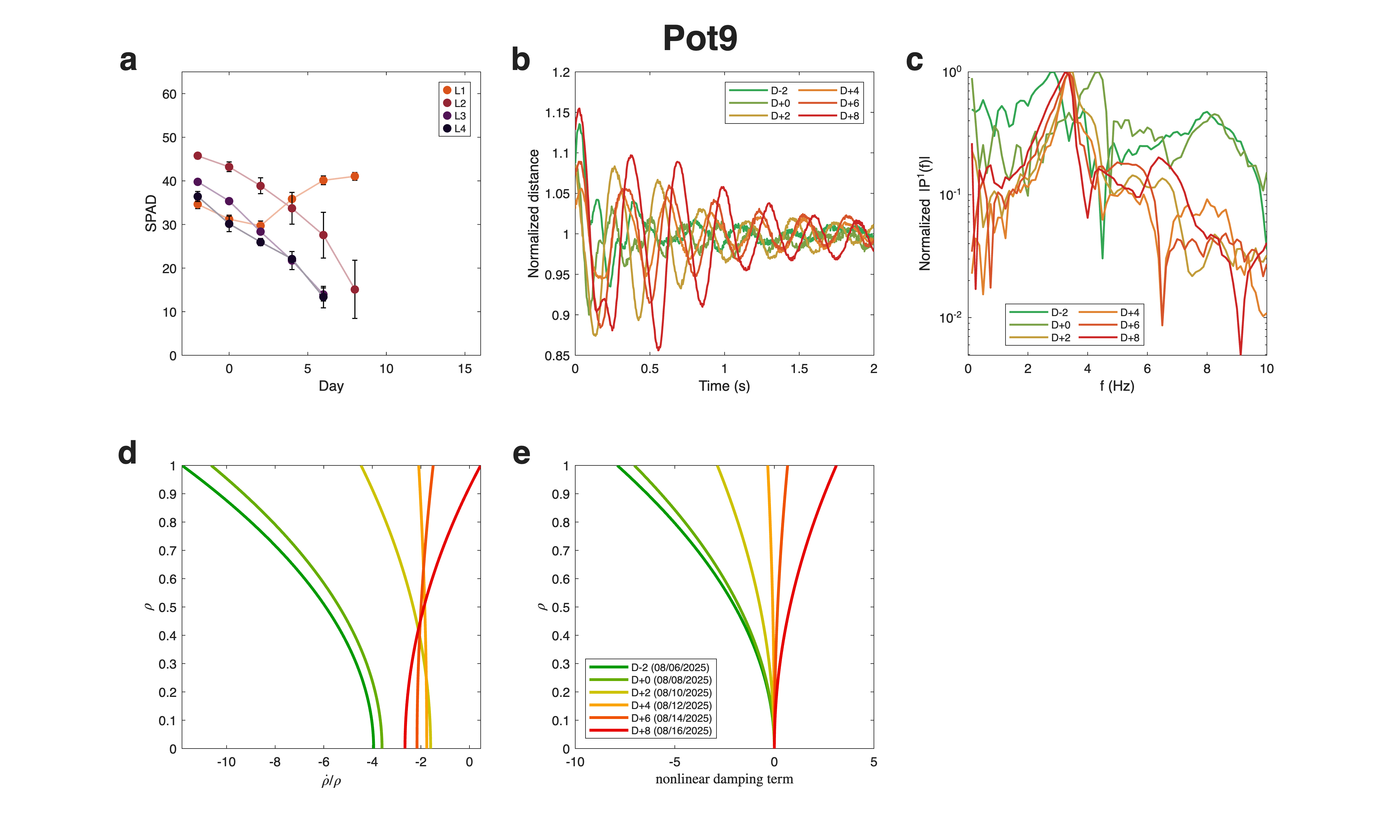}
    \caption{\textbf{Comprehensive experimental summary for Pot 9.} (a) SPAD values. (b) Time-series relative displacement. (c) FFT spectrum. (d) Total instantaneous damping rate ($\dot{\rho}/\rho$). (e) Nonlinear damping term. The loss of curvature in the damping profile mathematically reflects the loss of internal hydraulic pressure.}
    \label{FigS8}
\end{figure}

\clearpage

% --- SUMMARY FOR POT 10 ---
\begin{figure}[h]
    \centering
    \includegraphics[width=1\textwidth]{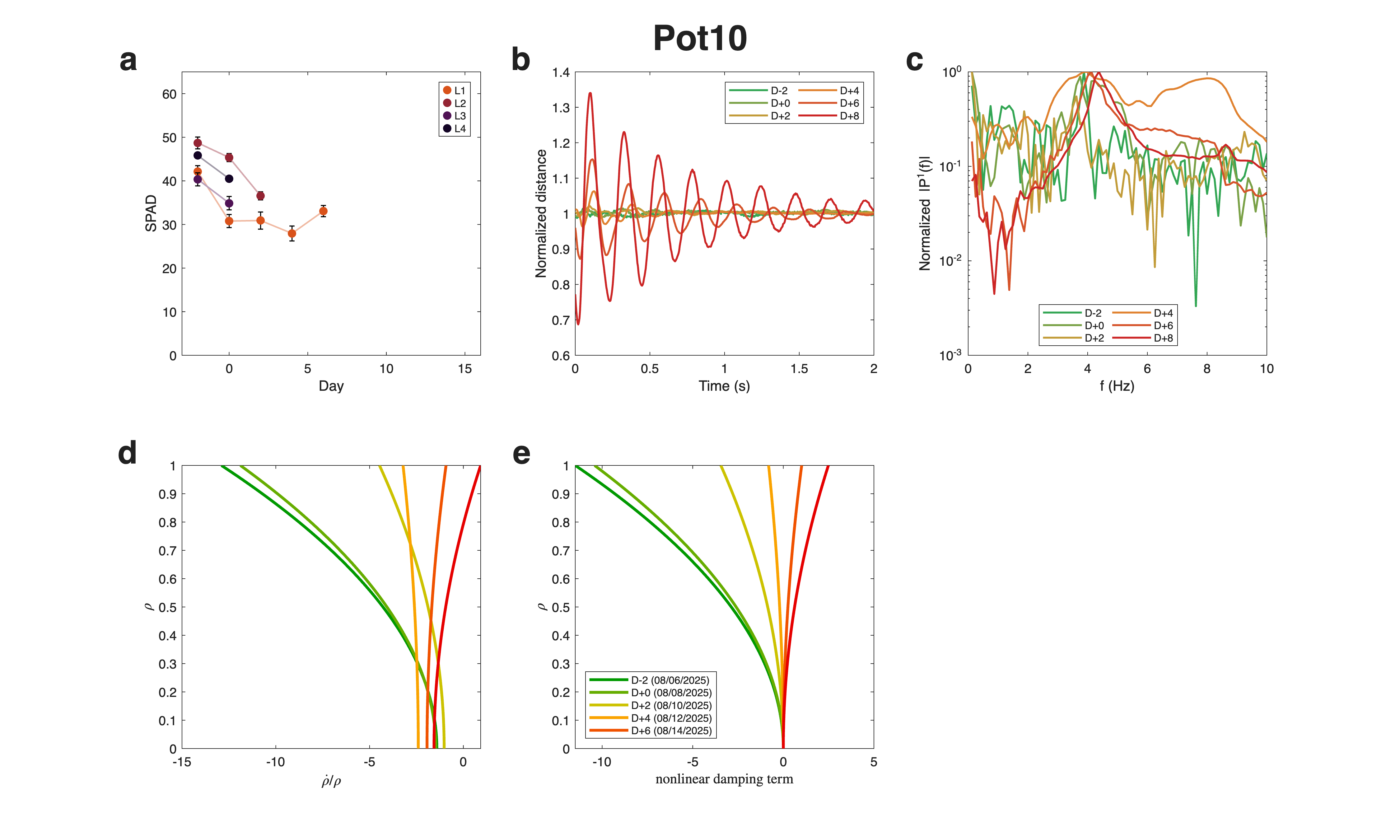}
    \caption{\textbf{Comprehensive experimental summary for Pot 10.} (a) SPAD values. (b) Time-series relative displacement. (c) FFT spectrum. (d) Total instantaneous damping rate ($\dot{\rho}/\rho$). (e) Nonlinear damping term. The individual plant dynamics confirm the universal trend of altered damping properties under severe water deficit.}
    \label{FigS9}
\end{figure}

\clearpage

% --- SUMMARY FOR POT 11 ---
\begin{figure}[h]
    \centering
    \includegraphics[width=1\textwidth]{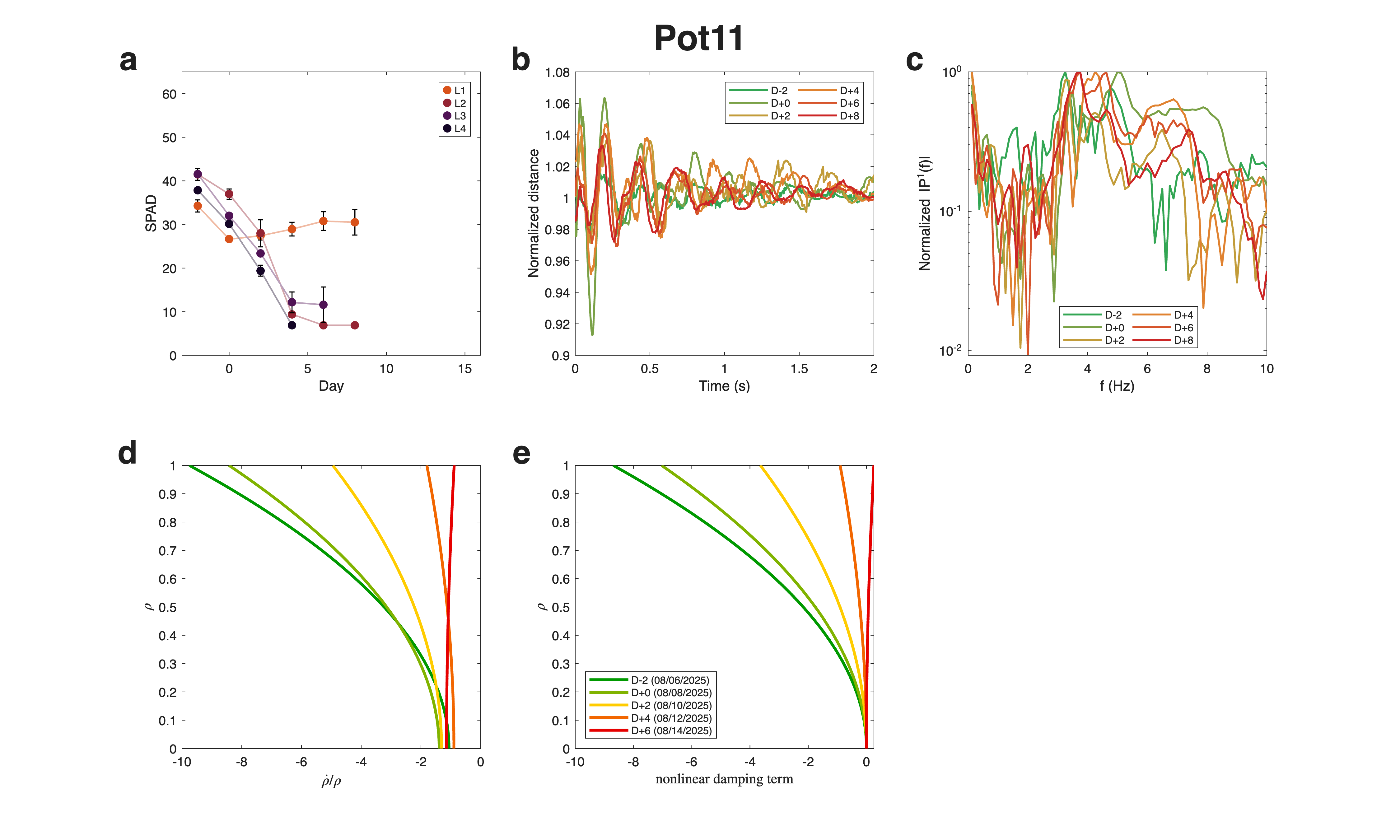}
    \caption{\textbf{Comprehensive experimental summary for Pot 11.} (a) SPAD values. (b) Time-series relative displacement. (c) FFT spectrum. (d) Total instantaneous damping rate ($\dot{\rho}/\rho$). (e) Nonlinear damping term. The mechanical shift to sub-linear damping correlates strongly with the late-stage physiological markers of structural dehydration.}
    \label{FigS10}
\end{figure}

% --- SUPPLEMENTARY FIGURE 1: Mechanical Sensitivity ---
\begin{figure}[h]
    \centering
    \includegraphics[width=0.75\textwidth]{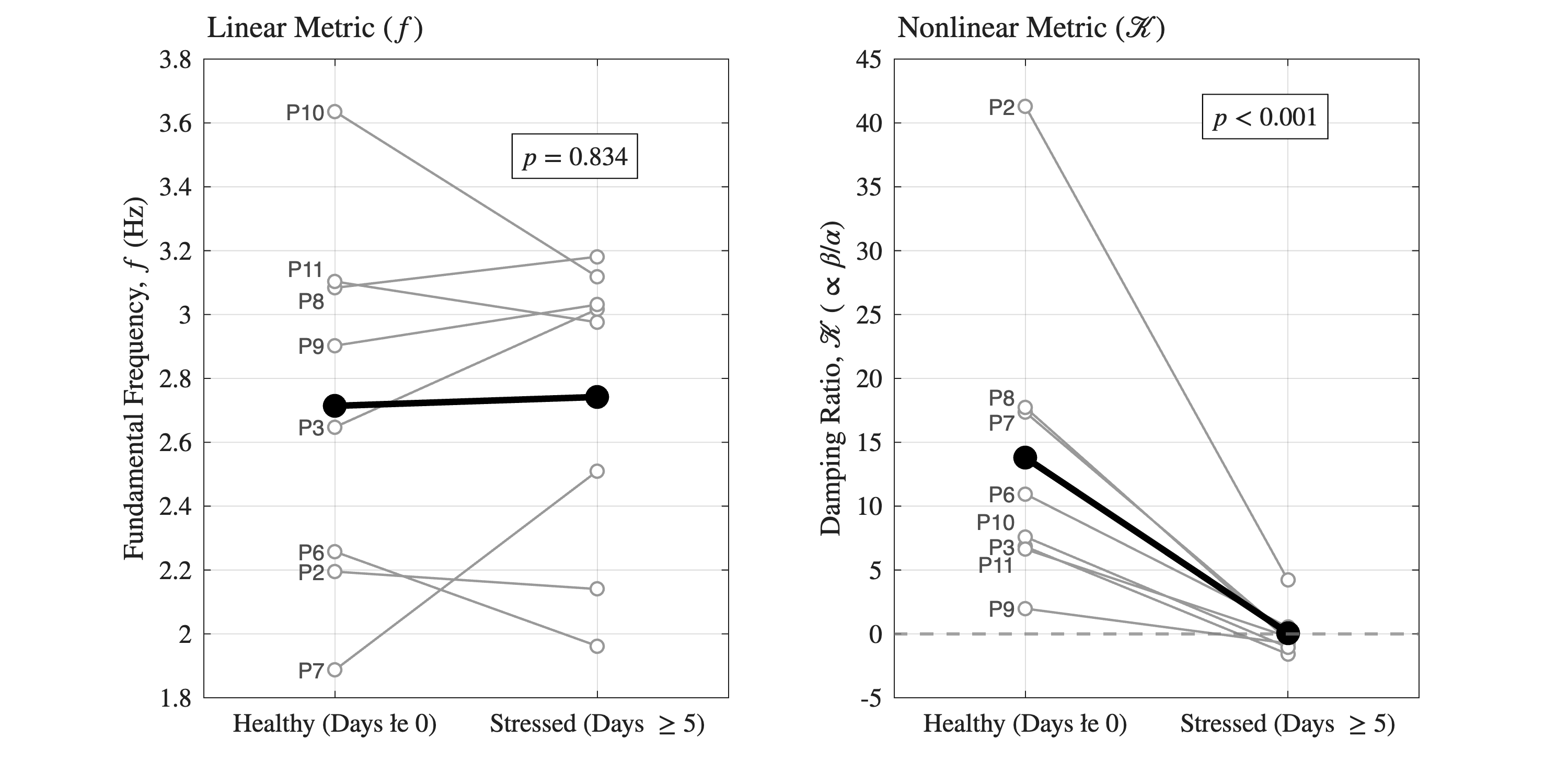}
    \caption{\textbf{Sensitivity of linear versus nonlinear mechanical metrics to severe drought stress.} 
    Data points represent individual plants (pots) evaluated at a healthy baseline (Days $\le 0$) and a drought-stressed state (Days $\ge 5$). The thick black line indicates the population mean trend. \textbf{(Left)} The linear fundamental frequency ($f$) exhibits no statistically significant shift during the drought period ($p = 0.834$), demonstrating high plant-to-plant variability and insensitivity to water status. \textbf{(Right)} Conversely, the dimensionless nonlinear damping ratio ($\mathcal{K} \propto \beta/\alpha$) demonstrates a highly significant, uniform drop across all tested plants ($p < 0.001$). Every plant universally crossed the mechanical bifurcation threshold ($\beta = 0$), transitioning from super-linear to sub-linear damping, proving $\mathcal{K}$ to be a highly robust, size-independent biomarker for drought stress.}
    \label{fig:mech_sensitivity}
\end{figure}

% --- SUPPLEMENTARY FIGURE 2: Physiological Sensitivity ---
\begin{figure}[h]
    \centering
    \includegraphics[width=\textwidth]{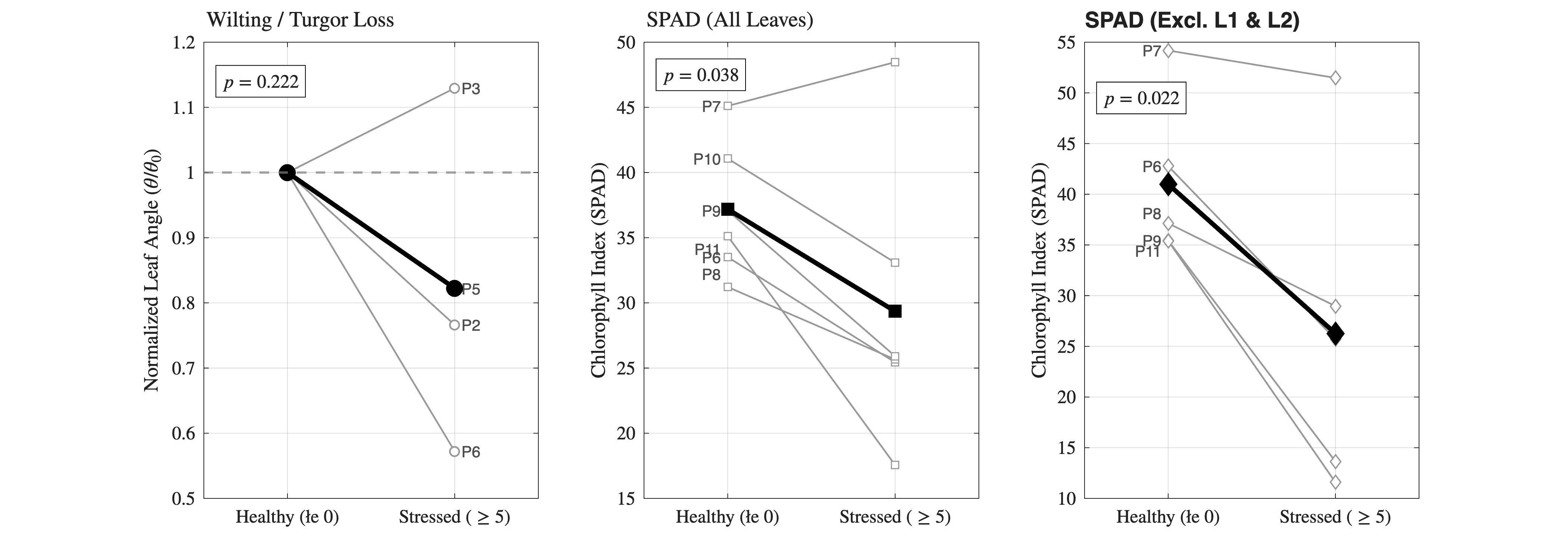}
    \caption{\textbf{Comparative sensitivity of physiological drought markers.} 
    Measurements tracking visual wilting and chlorophyll degradation from healthy (Days $\le 0$) to stressed (Days $\ge 5$) states. \textbf{(Left)} Normalized leaf angle ($\theta/\theta_0$) fails to provide a statistically significant drought indicator ($p = 0.222$) due to varying turgor loss strategies among individual plants. \textbf{(Middle)} Whole-plant chlorophyll index (SPAD) shows a significant degradation trend ($p = 0.038$). \textbf{(Right)} Excluding the mature lower leaves (L1 and L2)—which undergo early senescence—from the whole-plant average yields a much sharper, more statistically significant physiological signal ($p = 0.022$), confirming that the upper canopy dictates the primary survival response.}
    \label{fig:physio_sensitivity}
\end{figure}

% --- SUPPLEMENTARY FIGURE 3: Bifurcation Correlation ---
\begin{figure}[h]
    \centering
    \includegraphics[width=0.5\textwidth]{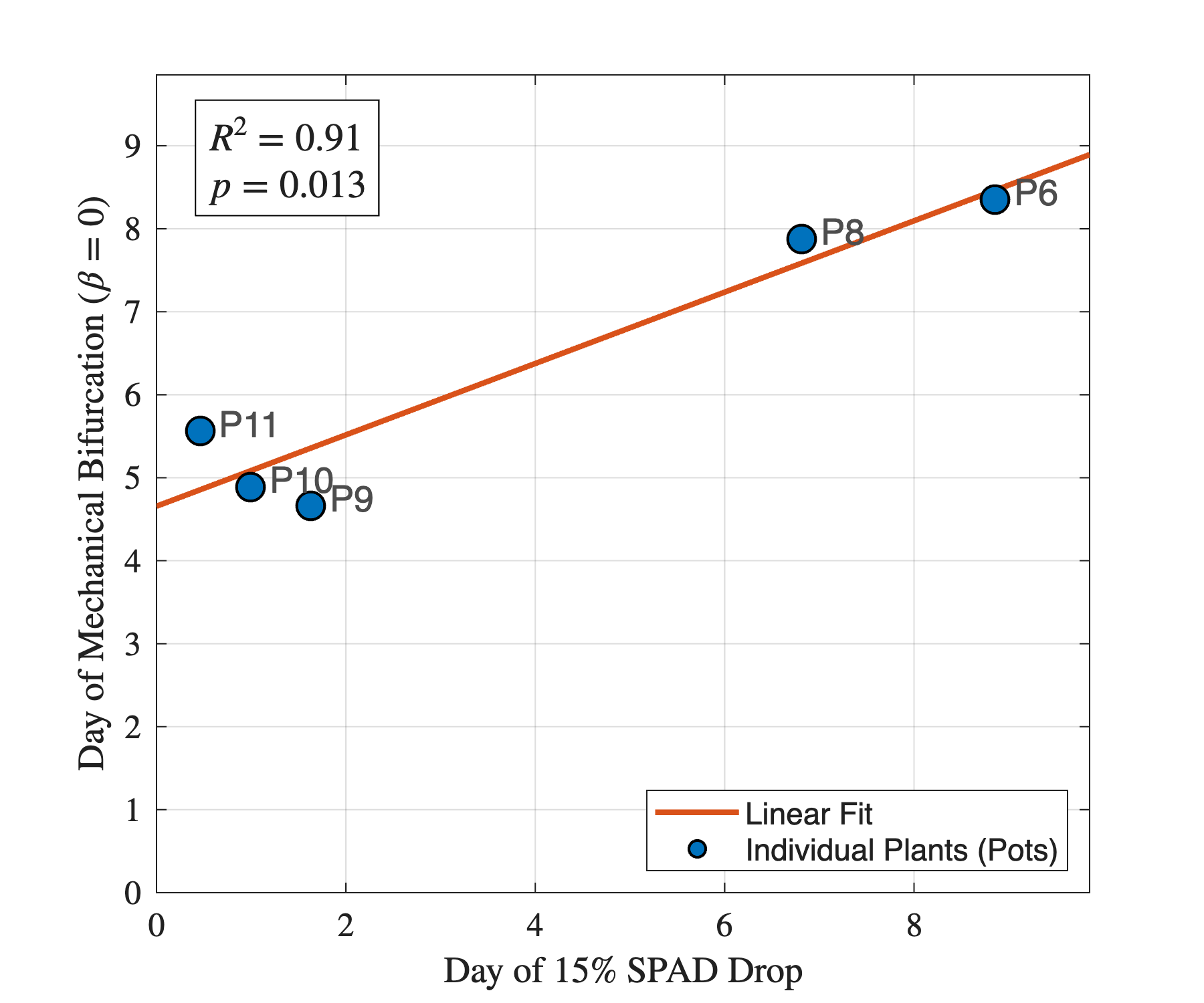}
    \caption{\textbf{Correlation between physiological degradation and mechanical bifurcation.} 
    Scatter plot mapping the day each plant experienced a 15\% drop in baseline SPAD (x-axis) against the exact day its stem mechanics transitioned from super-linear to sub-linear damping ($\beta = 0$, y-axis). The strong linear fit ($R^2 = 0.91$, $p = 0.013$) confirms that the nonlinear mechanical bifurcation is tightly coupled to severe internal physiological degradation. Plants that sustained their chlorophyll content longer (e.g., P6) correspondingly delayed their mechanical failure, validating the SSM-derived metric as a faithful proxy for biological drought resilience.}
    \label{fig:bifurcation_correlation}
\end{figure}

\clearpage
\section*{Supplementary Movie}

\noindent\textbf{Supplementary Movie~1: Free-vibration pluck tests and delay-coordinate phase portraits of a representative plant (Pot~6) under healthy and drought-stressed conditions.} \\
The upper panels show side-by-side high-speed recordings of the same plant before drought (Day~0, left) and after ten days of water stress (Day~10, right). Leaf landmarks are tracked frame-by-frame: the base (orange markers) and tip (green markers) of leaves L1--L3, with a yellow line joining each base to its tip. Following a pluck, the canopy sways from side to side with decaying amplitude. The lower panels show delay-coordinate phase portraits of the stem oscillation, reconstructed from a single scalar displacement signal $s(t)$: healthy (Day~$-2$, green, left) and drought-stressed (Day~10, red, right). A black marker traces the instantaneous state as each orbit is drawn. Together, the panels illustrate the plant's free-vibration response and its reconstructed dynamics, providing a visual comparison of healthy and drought-stressed conditions alongside the nonlinear damping changes identified by SSM reduction. The full clip is provided as \texttt{Movie\_Pot6\_Day0\_Day10\_LinkedIn.mp4}.